\documentclass{elsart}

\usepackage{psfig}
\newcommand{\beq}{\begin{equation}}
\newcommand{\eeq}[1]{\label{#1} \end{equation}}
\newcommand{\beqar}{\begin{eqnarray*}}
\newcommand{\eeqar}[1]{\label{#1} \end{eqnarray*}}

\newcommand{\insertplots}[1]{\centerline{\psfig{figure={#1},width=11.0cm}}}

\newcommand{\insertplotllarge}[1]{
      \centerline{\psfig{figure={#1},height=20.5cm}}
}
\newcommand{\insertplotlargee}[1]{
      \centerline{\psfig{figure={#1},height=23.0cm}}
}
\newcommand{\insertplotllargee}[1]{
      \centerline{\psfig{figure={#1},height=22.0cm}}
}
\newcommand{\insertplotlong}[1]{
      \centerline{\psfig{figure={#1},width=14.0cm}}
}

\newcommand{\dlt}{\bigtriangleup}

\begin{document}

\begin{frontmatter}



\title{Effective String Rope Model for the initial stages of Ultra-Relativistic
Heavy Ion Collisions}


\author[label1,label2]{V.K. Magas}
\ead{vladimir@cfif.ist.utl.pt}
\author[label2,label3]{L.P. Csernai}
\ead{csernai@fi.uib.no}
\author[label4]{D. Strottman}
\ead{dds@lanl.gov}

\address[label1]{
Center for Physics of Fundamental Interactions, Department of Physics\\
Instituto Superior Tecnico, Av. Rovisco Pais, 1049-001 Lisbon, Portugal
}
\address[label2]{
Section for Theoretical and Computational Physics, Department of Physics\\
University of Bergen, Allegaten 55, N-5007, Norway
}
\address[label3]{
KFKI Research Institute for Particle and Nuclear Physics\\
P.O.Box 49, 1525 Budapest, Hungary
}
\address[label4]{
Theoretical Division, Los Alamos National Laboratory\\
Los Alamos, NM, 87454, USA}

\begin{abstract}
Different approaches to describe initial stages of
relativistic heavy ion collisions are
discussed qualitatively and quantitatively.
An Effective String Rope Model is presented
for heavy ion collisions at RHIC energies. Our model takes into
account baryon recoil for both target and projectile, arising from the
acceleration of partons in an effective field, $F^{\mu \nu}$, produced in
the collision. The typical field strength (string tension) for RHIC
energies is about $5-12 \ GeV/fm$, what allows us to talk about ``string
ropes''.  The results show that a QGP forms a tilted disk, such that the
direction of the largest pressure gradient stays in the reaction plane,
but deviates from both the beam and the usual transverse flow directions.
The produced initial state can be used as
an initial condition for further hydrodynamical calculations.
Such initial conditions lead to the creation of third
flow component.
\end{abstract}


\end{frontmatter}

\section{Introduction}

\label{ch-1}

Fluid dynamical models are widely used to describe heavy ion collisions.
Their advantage is that one can vary flexibly the Equation of State
(EoS) of the matter and test its consequences on the reaction dynamics
and the outcome. This makes fluid dynamical models a very powerful tool
to study possible phase transitions in heavy ion collisions - such as
the liquid-gas or the Quark-Gluon Plasma (QGP) phase transition. For
example, the only models that can handle the supercooled QGP are
hydrodynamical models with a corresponding EoS. For highest energies 
achived nowadays at RHIC hydrodynamic calculations give a good 
description of the observed radial and elliptic flows 
\cite{Sollfrank-BigHydro,Schlei-BigHydro,Kolb-UU,Kolb-LowDensity,Kolb-Radial,Htoh}, in 
contrast to microscopic models, like HIJING \cite{Molnar-Elliptic} and UrQMD 
\cite{UrQMD-Elliptic}. 

In energetic collisions
of large heavy ions, especially if a QGP is formed in the collision,
one-fluid dynamics is a valid and good description for the intermediate
stages of the reaction. Here, interactions are strong and frequent, so
that other models, (e.g. transport models, string models, etc., that
assume binary collisions, with free propagation of constituents between
collisions) have limited validity. On the other hand, the initial and
final, Freeze Out (FO), stages of the reaction are outside the domain of
applicability of the fluid dynamical model.

Thus, the realistic, and detailed description of an energetic heavy ion
reaction requires a Multi Module Model, where the different stages of
the reaction are each described with a suitable theoretical approach. It
is important that these Modules are coupled to each other correctly: on
the interface, which is a three dimensional hypersurface in space-time
with normal $d\sigma^\mu$, all conservation laws should be satisfied,
and entropy should not decrease. These matching conditions were worked
out and studied for the matching at FO hypersurfaces  in details in Refs.
\cite{FO1,FO2,FO3}. In Ref \cite{antti} it has been shown that the 
entropy constraint is very sensitive to the pre-FO EoS, i.e.  the 
arbitrary choice of the pre-FO state does not necessarily lead to the  
physical FO. 

Similar ideas form the base of Ref. \cite{R32,Htoh,shurH} where authors 
combined two modules - hydro and UrQMD - 
replacing the hadronic phase of hydrodynamics with hadronic transport 
model to describe properly chemical and thermal FO.

The initial stages are the most problematic. We will discuss here
different models for the initial state in section \ref{ch-3} and will
present our modeling of it in section \ref{ch-4}.

Our goal is to build a Multi Module Model for ultra-relativistic heavy
ion collisions that is valid for RHIC and LHC energies, and maybe for
the most energetic SPS collisions. The present work is just the first
step, but an important one: the Effective String Rope model is
developed for the most problematic module - module describing the
initial stages of collisions.

\section{Initial
state of Relativistic Heavy Ion Collision}
\label{ch-3}
The situation at the beginning of an energetic heavy ion collision is
highly complicated and not easily understandable. None of the theoretical
models currently on the physics market can unambiguously describe the
initial stages.

As was pointed out in the introduction, at ultra-relativistic energies
the one-fluid dynamical models can not be justified for such a situation.
Frequently, two- or three-fluid models are used to remedy the difficulties
and to model the process of QGP formation and thermalization
\cite{A78,C82,bsd00}. Here the problem is transferred to the
determination of drag-, friction- and transfer- terms among the fluid
components, and a new problem is introduced with the (unjustified) use of
an EoS in each component in a nonequilibrated situation where an EoS is
not defined. Strictly speaking this approach can only be justified for
mixtures of noninteracting ideal gas components.

The kinetic transport models (see for example, Ultra-relativistic Quantum
Molecular Dynamics (UrQMD) \cite{BB98}) and parton cascades (see for
example, the Molnar Parton Cascade (MPC) \cite{MPC,MPC2}) are also
frequently applied to describe the high energy collisions from the very
beginning. Some of the parton cascades (for example, MPC) are able to
handle only $2\rightarrow 2$ reactions so far, what cannot be justified for
dense and strongly interacting systems.  The parton cascades also have
to approximate media-effects, since the parton structure function,
measured for a free nucleon, is modified inside the nucleus. This is the
so-called shadowing effect \cite{EMC}.  This approximation procedure is
also not unambiguous, see for example Ref. \cite{GW94,shadow} for more
details.

Some models, like UrQMD, were initially developed for much  lower
energies than available nowadays, but even then, when the particle
propagates in a hot and dense and  nonequilibrated medium of highly
excited hadrons, the properties of  this particle might change
significantly. Thus, properties like  effective masses, decay widths,
effective momenta, and in-medium cross sections have to be calculated for
a given local situation at the point where particle propagates.
Unfortunately this is a very complicated and numerically very time
consuming task, so in most of the models drastic approximations are
made (see for example \cite{BB98}).  In a relativistic heavy ion
collision, especially at the early stages, the excitation energy per
particle would be so high that the concept of resonance (we have to
deal not only with nucleons, but with many different baryons and mesons)
has no sense any more. Here the quark and gluon degrees of freedom must
be taken into account. Therefore, some models, including UrQMD, replace
the resonances by continuous string excitations for internal excitation
energies higher than approximately $2\ GeV$. The properties of such
effective ``strings'' are basically unknown and can be varied within a wide
range, almost like free parameters \cite{Goren}.

Perturbative Quantum Chromo Dynamics (pQCD) would be, in principle, the
proper model for describing our systems at very high energies.
Unfortunately, pQCD itself is not applicable for heavy ion reactions at
RHIC energies. Nevertheless, the pQCD calculations with some extra
non-perturbative, phenomenological assumptions, like saturation of a
gluon plasma, can be performed. Different models following this scenario
have been proposed in \cite{McV94,muell87,EK99,EK,EK00,EK01-1}. The
extension of the pQCD+saturation model for nonzero impact parameters,
which is of primary interest for us, leads to serious complications in
calculations \cite{EK00}.

Thus, since we can not unambiguously describe the initial stages of
relativistic heavy ion collisions in microscopical models, we have to
use some phenomenological models (in fact all the microscopical models
discussed above already have some phenomenological elements).

\subsection{Different models}
\label{dm}

There are two phenomenological models, which are most frequently
discussed in the literature, and which claim to give us at least a qualitative
understanding of the initial stages of the relativistic heavy ion
collisions.

\subsection{The Landau model}
\label{lm}

The Landau model has been developed to generalize the fluid dynamical
approach to the energy range where we can not neglect relativistic
effects, such as Lorentz contraction of the colliding nuclei. Actually
the model was originally developed for $pp$ collisions, but the
generalization for $AA$ is rather straightforward and the fluid dynamical
expansion is even more justified for bigger systems.

\begin{figure*}[htb]
\insertplotlong{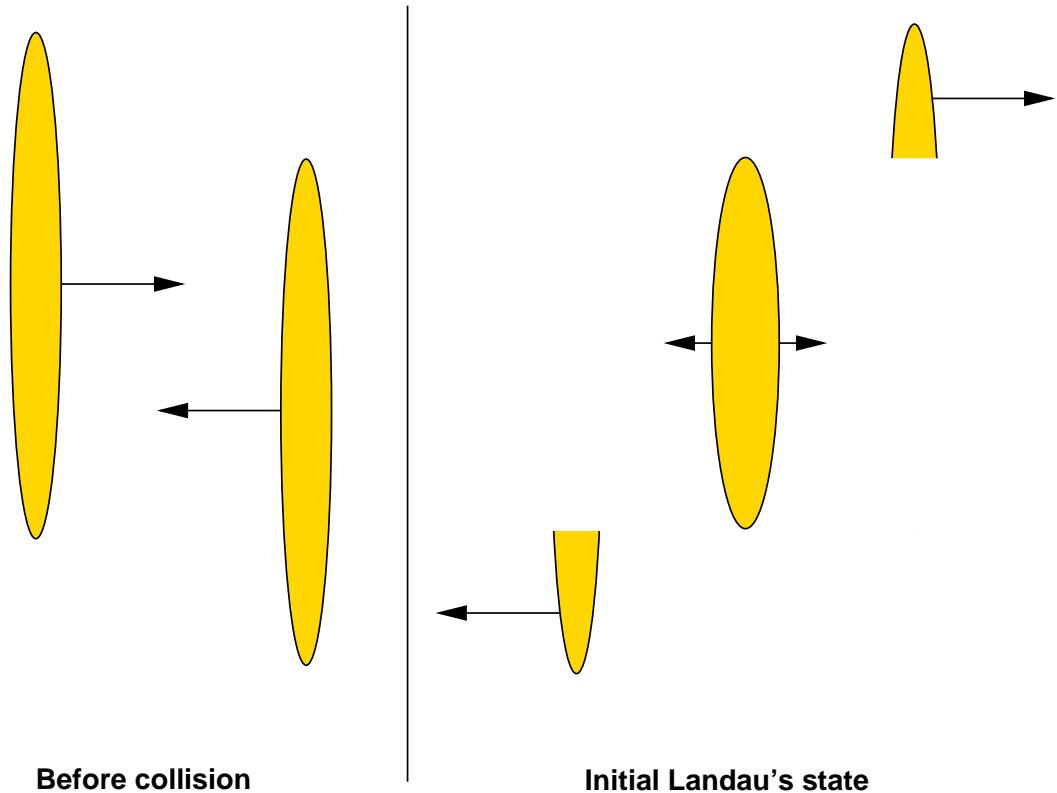}
\caption[]{Initial state in Landau model.}
\label{lanf}
\end{figure*}

It is the most appropriate model for relativistic heavy ion collisions in
the energy range $E_{lab}=10-100\ A\cdot GeV$. For energies above
$\gamma^{CM}\approx 3-10$, so in the Center of Mass (CM) frame the
colliding nuclei are considerably contracted and have disk-like shapes.
The Landau fluid dynamical model assumes the initial thermalized state as
a static, completely stopped, homogeneous disk, contracted in the beam
direction by the factor $\gamma_{CM}$ (see Fig. \ref{lanf}). The fluid
dynamical expansion initially starts in the direction of the largest
pressure gradient - in the beam direction in our case. Only when the
system has expanded in the $z$ direction to a size comparable to its
transverse diameter the transverse expansion also considered. In this
second phase the expansion is 2+1 dimensional, while initially it was 1+1
dimensional (all quantities were $z$ and $t$ dependent).

The inputs to this model are: A) the equation of state; B) the initial
energy and  baryon densities; C) the FO conditions.  Usually the initial
energy density, $e_0$, is estimated in the following way:
\beq e_0=E_{CM}/V_0\ , \quad {\rm where} \quad V_0=V_{rest}/\gamma_{CM}\ .
\eeq{ine_lan}

\subsection{The Bjorken model}
\label{bm}

If we increase the collision energy further to $E_{CM}=100\ A\cdot GeV$
or more, the expectation is that the concept of complete stopping does not
work anymore. Furthermore, when two nucleons from the target and
the projectile nuclei collide at these energies, their momenta are so high
that they see each other as compound objects, containing three valence
quarks and other partons. The point is that the proper participants at
these energies are the partons and not the nucleons. The initial momenta
of partons are very high and we expect these two parton
clouds to cross each other and keep going with almost their original
rapidities. Thus, contrary to complete stopping assumed in the Landau
model,  we have now the opposite scenario
 - nuclei become essentially transparent to
each other.

During their interpenetration, however, the target and projectile partons
may exchange color charges, what leads to the creation of a chromoelectric
field in the middle of the reaction volume. The classical analogy, that one
can keep in mind for a better understanding, is the electric field between
two condenser plates, moving away from each other. From QCD calculations
we know that such a chromoelectric field should be confined in the
transverse direction and have the geometry of a string. The energy density
of such a string is substantial - already for the strings produced in
hadronic collisions the string tension is about $1\ GeV/fm$ , and the
theoretical simulations show that even more energetic objects may be
produced in heavy ion collisions. Later these strings start to decay by
producing quarks, antiquarks 
and gluons, but because of baryon number conservation,
the matter in the middle of reaction volume is baryon free. Theoreticians
like very much this last feature of the model, since the baryon free
matter is rather pleasant for calculations.

\begin{figure*}[htb]
\vspace{0.6cm}
\insertplotlong{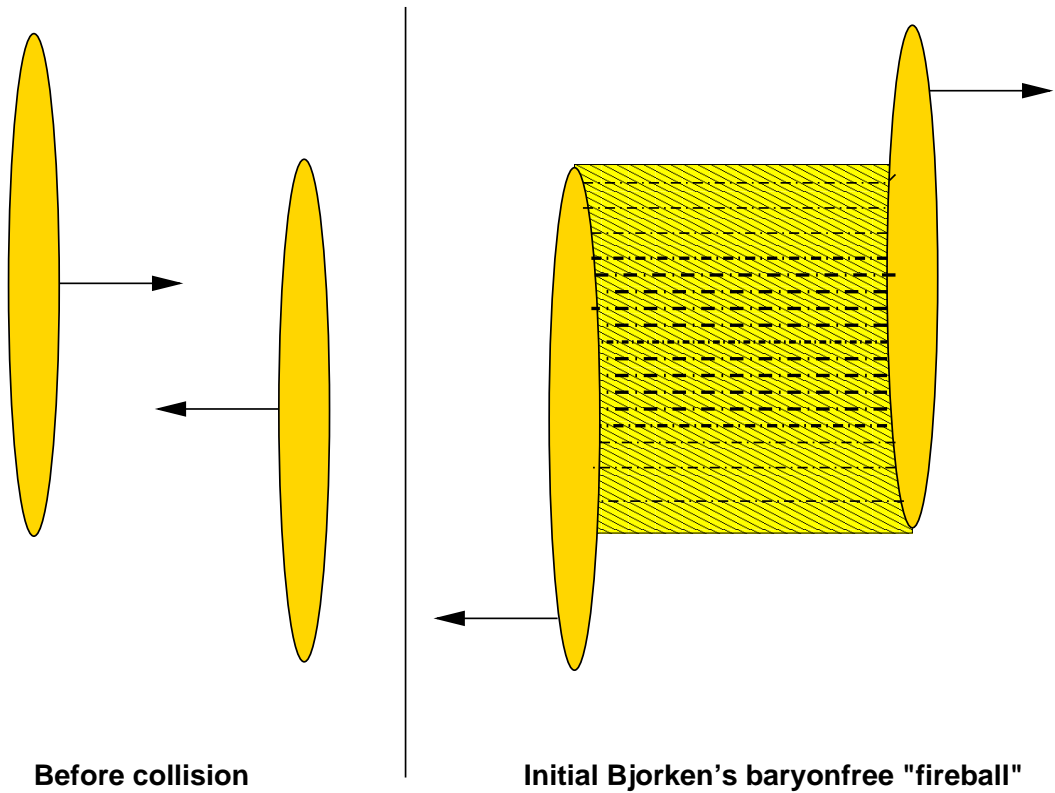}
\caption[]{Initial state in the Bjorken model. The dash-dotted lines denote the
chromoelectric field strings.}
\label{bjorf}
\end{figure*}

The important input to the model is given by experimental data from
ultra-relativistic $pp$ collisions. The rapidity distribution of the
charged particle multiplicity, $dN_{ch}/dy$, can be divided, with
reasonable accuracy, into three regions. Two smooth peaks around $y=\pm
y_0$, where $y_0$ is the original rapidity, are believed to be originated
by target and projectile partons, interpenetrating through each other
almost without rapidity lost. Between those we can see the plateau, -
$dN_{ch}/dy=const$, - generated by our baryon free flux tube. Assuming the
same symmetry, $dQ/dy=const$, for all other quantities a simple
hydrodynamical model with an analytical solution has been developed by
Bjorken in Ref. \cite{bjor}. The above assumption, $dQ/dy=const$, means
that our model is boost invariant - all the quantities are functions of
the proper time only, $Q=Q(\tau)$, where $\tau=\sqrt{t^2-z^2}$. Obviously
the model has a 1+1 dimensional structure.

In order to study transverse expansions, the Bjorken model has to be
supplemented with a description of this process separately, see for
example \cite{las22}. The boost invariance of the model is also very much
appreciated by theoreticians, since it is then sufficient to perform
simulations for the $y=0\ (z=0)$ plane only, and then boost the result to
reproduce all of the phase space. The Bjorken model does not tell us
anything about the energy density distribution in the plane perpendicular
to the beam direction, say the $z=0\ (y=0)$ plane; therefore additional
assumptions are required. If we define a grid in the $[x,y]$ plane, then
the energy density profiles proportional to the number of binary
collisions,
$e(x_i,y_j,z=0)\propto N_{target}(x_i,y_j)\cdot N_{target}(x_i,y_j)$, or
proportional to number of participants,
$e(x_i,y_j,z=0)\propto \left(N_{target}(x_i,y_j)+
N_{target}(x_i,y_j)\right)$, are the most popular assumptions in literature
(see
\cite{EK01-p} for recent overview).

The initial energy density in the Bjorken model can be estimated in the
following way. At the $y=0\ (z=0)$ plane the full energy is the
transverse energy, $E_{full}=E_{T}$. One may also notice that for the
Bjorken boost invariant model, the pseudo-rapidity 
($\eta= - \ln{\tan{\frac{\theta}{2}}}$), 
where $\theta$ is a polar angle of emitted particle, 
is the same as rapidity, $\eta=y$. If in the
experiment  we measure the  energy $\dlt E_{T}$ emitted in a very small
rapidity window, $[0,\dlt y]$, then  we can estimate the initial energy
density:
$$ e_0=\frac{\dlt E_{full} }{\dlt V}=
\left(\frac{\dlt E_{T}}{A \dlt z }\right)_{y=0}\ .
$$ Now using the following expressions: transverse area $A=\pi R_{A}^2$,
where $R_A$ is the  radius of nucleus with A nucleons,  and size in the
beam direction
$\dlt z =\tau \dlt \eta = \tau_0 \dlt y$, where $\tau_0$ is
the thermalization time, -  we obtain
\beq
e_0=\frac{1}{ \pi R_{A}^2 \tau_0}\left(\frac{d E_{T}}{d y}\right)_{y=0}\ .
\eeq{ine_bjor}
Here the thermalization time is unknown and should be estimated from some
other model for the particular reaction (for example, $\tau_0\approx 1\ fm$
is usually taken for SPS Pb+Pb $158\ GeV/nucl$ collisions). Another question
is - can we really use the post FO final state measurement,
$\left(\frac{d E_{T}}{d y}\right)_{y=0}$, directly to estimate the initial
energy density.  Briefly speaking, this means to  neglect the longitudinal 
expansion
caused by pressure during the evolution of our fireball,  which would
reduce our initial energy density to the one estimated from  the data by
eq. (\ref{ine_bjor}). Therefore, one may expect that the initial energy
density is larger than the one given by this simple estimate.

\subsection{Bjorken model with baryon recoil}
\label{bmw}

Certainly, the assumption of non-stopping partons is an idealization:
from basic physical principles we may expect that if two moving quarks
stretch a string between themselves with a substantial string tension,
they should be decelerated, since during such a process they are
converting their kinetic energy into the energy of the string. Thus, the
creation of the chromoelectric field should produce some baryon recoil for
both colliding nuclei. In the first approximation this was done in Ref.
\cite{GC86}.

The authors developed the model for $pA$ collisions. The initial momentum
of the projectile, $p$, was considered to be so huge, that the recoil
effect could be disregarded for it, while for the target partons the
recoil arises from the acceleration of partons in an effective external
field $F^{\mu\nu}$, produced in the collision. This baryon recoil was
included in a way that guarantees the conservation of baryon current. At
the same time the energy-momentum conservation was obviously violated -
the effective field produced in the collision was considered to be
external, and the projectile matter was not taken into account in the
energy-momentum tensor of the system. Thus, the authors ended up with the
following system of the modified fluid dynamical equations:
\beq
\partial_\mu T^{\mu\nu}=F^{\nu\mu} N_\mu + \Sigma^\nu_\pi\ ,
\eeq{rec1}
\beq
\partial_\mu N^\mu = 0 \ ,
\eeq{rec2}
where
$
T^{\mu\nu}=e_t\left(\left(1+c_0^2\right)u_t^\mu u_t^\nu
- c_0^2g^{\mu\nu}\right)$
is an energy-momentum tensor (only target matter is taken into account),
$c_0^2$ comes from the EoS $P=c_0^2 e$,
$\Sigma^\nu_\pi$ is the pion source term \cite{las22}, and $N^\mu = n
u^\mu$.  The effective field, $F^{\mu\nu}$, is parametrized in the
following way:
\beq
F^{\mu\nu}=\left(
\begin{array}{cccc}
0 & 0& 0& -\sigma \\
0 & 0& 0& 0\\
0 & 0& 0& 0\\
\sigma& 0& 0 & 0
\end{array}\right)
  \ \ ,
\eeq{eff0}
where
$\sigma = const$ is the field strength, or string tension, of the above
described effective external field. The authors have found
$\sigma \approx 3-4\ GeV/fm$.

Without the source term the system (\ref{rec1}, \ref{rec2}) can be solved
analytically \cite{GC86}. The source term is necessary at the later stages
of the reaction, since it allows the neutralization process, i.e. the
mechanism by which the energy stored in the field is converted back into
the energy of matter (see \cite{GC86} for details). The interplay between
the baryon recoil and the energy, coming from the neutralization of the
strings - string decay,  leads to much higher energy density in the
fragmentation region  than the original Bjorken model predicts. We will
come up with the modification of this  nice simple model in section
\ref{ch-4}.

Experiments have now entered the region where the Bjorken model is
expected to be applicable, but there is no clear and unambiguous
confirmation that Nature follows this scenario. The preliminary
experimental results from RHIC do not show transparency - most particle
multiplicities do not show a dip in the rapidity spectra, but rather a
plateau around mid-rapidity\cite{QM01,multipl}, which is a sign of strong
stopping. Furthermore, a very strong elliptic flow ($v_2$ flow
component\footnote{Nowadays the expansion in Fourier series is usually
applied to study the azimuthal distribution of particles
\cite{VoZh96,PoVo98}:
$$ E \frac{d^3 N}{d^3 p} = \frac{1}{\pi} \frac{d^2 N}{dp_t^2 dy} \left[
1 + \sum_{n=1}^{\infty} 2 v_n \cos(n\phi) \right] ,
$$ where $\phi$ is the azimuthal angle. The first term in square brackets
represents the isotropic radial flow, while the others are referred to
anisotropic flow. The first Fourier coefficient $v_1 = \langle \cos{\phi}
\rangle \equiv \langle p_x / p_t \rangle $ is called directed flow, and the
second one
$v_2 = \langle \cos{(2 \phi)} \rangle \equiv \langle (p_x / p_t)^2 - (p_y
/ p_t)^2 \rangle $ is called elliptic flow. }) has been measured, which
shows a clear peak around mid rapidity \cite{QM01,v2}. To build such a
strong elliptic flow, strong stopping and momentum equilibration are
required. Also the $\bar{p}/p$ ratio at mid-rapidity measured at RHIC
\cite{QM-vid,prstar} (preliminary) is still far from one, which tells us
that the middle region is not baryon-free.

\subsection{Third flow component}
\label{3flow}

As discussed in the introduction, the fluid dynamics is governed by the
Equation of State of the matter, and the analysis of the resulting flow
patterns turned out to be one of the best tools to extract the EoS from
the outcome of a heavy ion collision. The final event shape must carry
the information about the pressure development during the collision
including the early stages of the collision.

The phase transition to the QGP is connected to a decrease of pressure
according to most theoretical estimates, not only in strong first order
phase transition models, but even if we have a smooth but rapid, gradual
transition. This reduced pressure  around the phase transition threshold
are known for long  (see e.g. \cite{holme}), and it was emphasized as a
possible QGP signal:  this ``soft point'' of the EoS might be possible to
observe in excitation functions of collective flow data
\cite{rpm95,shur95}. It was pointed out earlier \cite{hpg93} that the
decrease in the out-of-plane (squeeze out) emission is even more sensitive
to the pressure drop and it decreases due to plasma formation stronger
than the in-plane collective flow.

In this section we want to discuss another consequence of the same
softening in the EoS, which is a recently (1999) identified new distinct
flow pattern - the third flow component \cite{CR}. This flow pattern can
be seen in almost all theoretical fluid dynamics calculations with QGP
formation, and also in experimental data, but it was not discussed
earlier, see for example \cite{hpg93,ASC91prl,bcl94,rpm95,bdm97}.

Two basic flow patterns have been predicted and detected before at lower
energies: \\ A) the directed transverse flow in the reaction plane, or
side-splash, or bounce off, which is most frequently presented on the well
known $p_x$ vs. $y$ diagram (or $v_1$ flow component vs. $y$)
  and seen at all energies in heavy ion collisions from energies of
$30\ A\cdot MeV$ to $165\ A\cdot GeV$ \cite{e877,liu97,na98,wa98}; \\ B)
the squeeze-out effect which is an enhanced emission of particles
transverse to the reaction plane at center-of-mass rapidities.

At lower energies the directed transverse flow resulted in a smooth,
linear $p_x$ vs. $y$ dependence at CM rapidities. This straight line
behavior connecting the maximum at $y_{proj}$ and the minimum at $y_{targ}$
was so typical that it was used to compare flow data at different beam
energies and masses.

At higher energies, say above $10-11\ A\cdot GeV$ in Lab frame, the
deviations from this straight line behavior have been observed
\cite{lpx98,e877,na98,wa98}. Unfortunately, directed flow has not yet
been measured at RHIC.

In theoretical calculations the important feature of this third flow
component is that it is clearly seen in fluid dynamical calculations when,
and only when, the QGP formation was allowed during the calculation (see
\cite{CR} for overview). In sharp contrast, the solutions with a purely
hadronic EoS did not show this effect, and the maximum and minimum of the
$p_x$ curve could be connected with a straight line\footnote{Note that all
these fluid dynamical calculations were done much before the experiments.
The first quantitative flow predictions \cite{ASC91prl} preceded the
experiments by six years and gave rather good agreement with the data.}.

\subsection{Possible source of the third flow component in fluid dynamics}
\label{pso}

\begin{figure*}[htb]
\insertplotlong{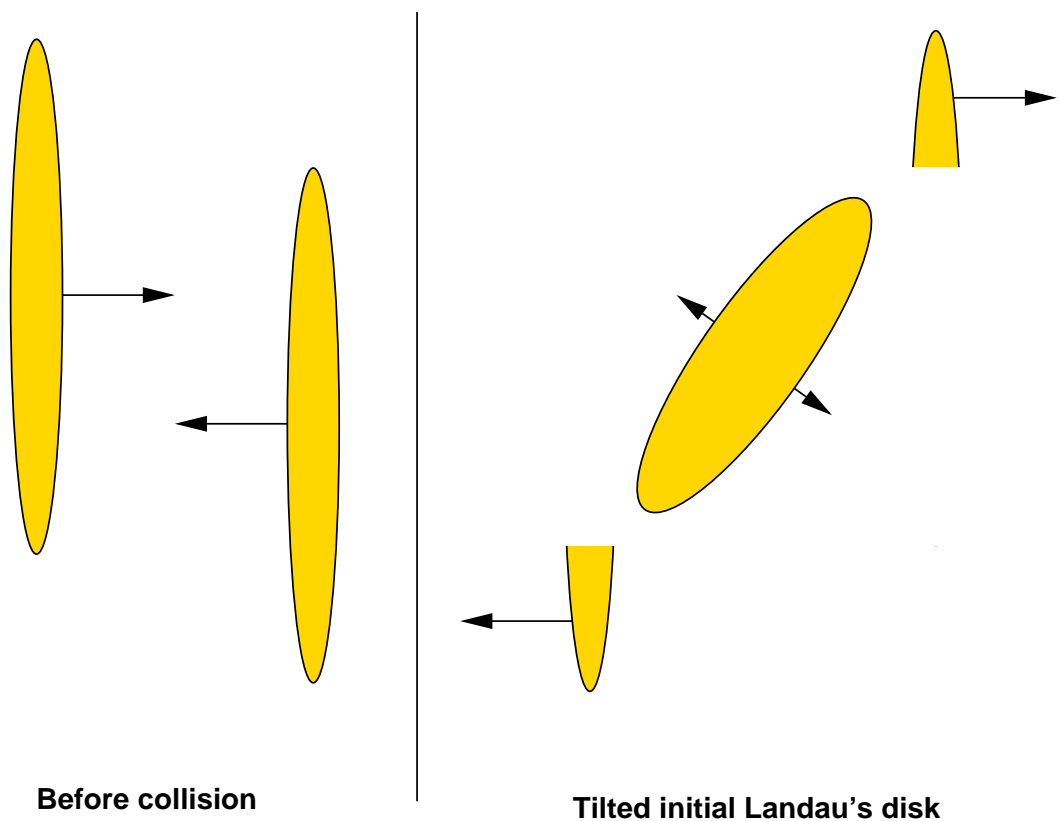}
\vspace{0.3cm}
\caption[]{Tilted initial state. The direction of the largest pressure
gradient stays in the reaction plane, but deviates from both the beam and
the usual transverse flow directions. Such initial conditions may lead to
the creation of the third flow component \cite{CR}.}
\label{tilf}
\end{figure*}

The following possible source of the third flow component has been
proposed in Ref. \cite{CR}.
If QGP is formed in the collision, strong and rapid stopping may take
place, and a behavior close to that of a one-fluid model one might be
established. I.e., let's assume that Bjorken scenario is not realized, at
least in its form discussed in previous section, but that the Landau fluid
dynamical model is applicable for collisions of massive ultra-relativistic
heavy ions (see the discussion at the and of section \ref{bmw}). The soft
and compressible QGP forms a rather flat disk which is at rest in the CM
system (see Fig. \ref{tilf}). Then this disk starts to expand in the
direction of the largest pressure gradient. At small but finite impact
parameters we may assume that this disk is tilted, and the direction of
fastest expansion will stay in the reaction plane, but will deviate from
the both beam axis and the usual transverse flow direction. Thus, a third
flow component might develop from the large pressure gradient and tilted
and strongly Lorentz contracted initial state. Now the flow in the
reaction plane will have an ``elliptic structure'' \cite{na98,pw98,so99}
and must be characterized by two flow components: the usual transverse
flow and the one orthogonal to it - the third flow component or antiflow.
At the same time as the primary directed flow is weakened by the stronger
Lorentz contraction at higher energies, this third flow component is
strengthened by increased pressure gradient arising from the Lorentz
contraction.

\begin{figure*}[htb]
\vspace{0.6cm}
\insertplotlong{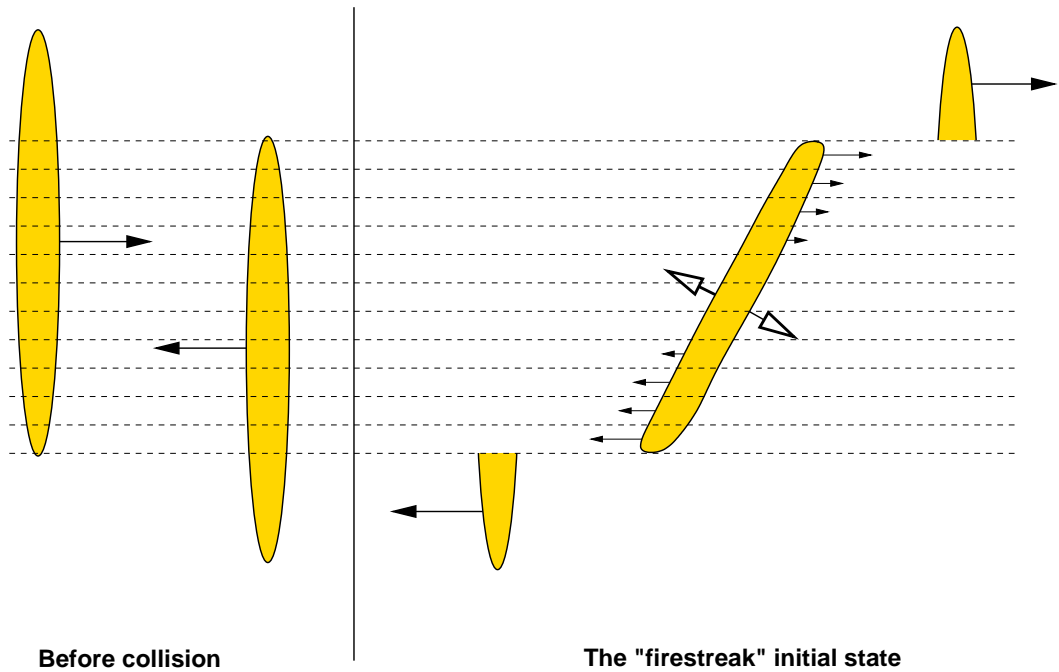}
\caption[]{The ``firestreak'' initial state.}
\label{fs}
\end{figure*}

The concept of a tilted ``fireball'' was proposed initially in the so-called
``firestreak'' models \cite{frstr1,frstr2}. The idea behind it is rather
simple and has to do with momentum conservation. Without going into
mathematical details we can understand it in the following way. First, let
us create a grid in the $[x,y]$-plane and subdivide the projectile and
target into streaks parallel to the beam direction. We try to describe
the collision with finite impact parameter. In the CM frame in one
streak the pieces of matter from target and projectile will have the same
absolute values of rapidities, but different momenta, since they in
general contain different amount of matter (see Fig. \ref{fs}). If the
collision happens so fast that there is no interaction between streaks,
then the momentum conservation does allow complete stopping, i.e., the
final streak - the piece of matter within one tube with the same
transverse coordinates after collision of target and projectile fragments
- will have a momentum defined by momentum conservation, and
correspondingly its center of mass has to move with some rapidity.
Crossing the participant region along $x$ axes in the reaction plane we
will see that in the middle the final streaks rapidities are small, since
the target and projectile fragments were almost equal; while at the edges
of the participant region the final streaks move rather fast (see Fig.
\ref{fs}). Thus, after some time our tilted ``fireball'' will degenerate
into a ``firestreak''. Please notice the difference between ``firestreak''
and the tilted Landau disk discussed above: the tilted disk was at rest in
CM frame, i.e., the complete stopping has been assumed, while ``firestreak''
geometry is dynamic. By ``dynamic'' we mean that different parts of
``firestreak'' move with different rapidities.

In ``firestreak'' models such a geometry has been assumed at the freeze-out
time, and then, based on
energy, charge and baryon number conservations (we already conserve
momentum) the authors tried to reproduce the parameters of the post FO
distribution \cite{frstr2}. This procedure is somewhat similar to the
statistical production model, which is used nowadays to calculate
production of different hadronic species at chemical FO hypersurface. Such
calculations were not really supported by experiment. Nevertheless, in
our opinion, this scenario might be a good qualitative description for
the initial stages of the heavy ion collisions until the local
thermalization achieved. When the matter is locally thermalized and the
pressure is built up, we cannot neglect hydrodynamical expansion, which
will smear out this initial distribution producing a more or less
spherical fireball before the freeze out process starts. Nevertheless,
such a hydrodynamical expansion will initially start in the direction of
the largest pressure gradient (see Fig. \ref{fs}) and, together with the
initial velocity field, might produce a two component (elliptic) structure
of the flow in the reaction plane, identified in ref. \cite{CR}.

\section{Effective String Rope model}

\label{ch-4}

In this section we will discuss so called Effective String Rope model 
(ESRM) for
energy, pressure and flow velocity distributions at the beginning of
ultra-relativistic heavy ion collisions. The output of this model can  be
used as an initial condition for further hydrodynamic calculations.

One important conclusion of heavy ion research in the last decade is that
standard 'hadronic' string models fail to describe heavy ion experiments.
All string models had to introduce new, energetic objects: string ropes
\cite{bnk84,S95}, quark clusters \cite{WA96}, or fused strings
\cite{ABP93}, in order to describe the abundant formation of massive
particles like strange antibaryons. Based on this, we describe the initial
moments of the reaction in the framework of classical, or coherent,
Yang-Mills theory, following the Bjorken model with baryon recoil  Ref.
\cite{GC86} (see section \ref{bmw}) assuming a larger field strength
(string tension) than in ordinary hadron-hadron collisions. For example,
calculations both in the Quark Gluon String Model (QGSM)
\cite{ASC91pl,ASC91prl,ACS92}
and in the Monte Carlo string fusion model \cite{ABP93} indicate that the
energy density of strings reaches $8\ -\ 10\ GeV/fm$ already in SPS
reactions. This is nearly $10$ times more than  the tension used in
standard, 'hadronic', string models where $\sigma \approx 1\ GeV/fm$.  In
addition we now satisfy all conservation laws exactly, while in Ref.
\cite{GC86}, as it was discussed in section \ref{bmw}, energy-momentum
conservation was irrelevant and was ignored.  Thus, in this approach for
the first time the initial transparency/stopping and energy deposited into
strings and ``string ropes'' was determined consistently with each
other \cite{MCS01}.

We do not describe the hadronization of our strings. Thus, this approach can
be justified only for the very initial stages. The effective way to perform
string decay will be discussed in section \ref{four}.

Recent parton kinetic models indicate that saturation of a gluon plasma
takes place in a very short time - $\tau_{sat}=0.09,0.27\ fm/c$ for LHC,
SPS energies respectively \cite{EK99}, and a pressure needs $\tau_p=1-2\
fm/c$ (SPS) and $\tau_p=3-5\ fm/c$ (LHC) to be established \cite{DG00}.
More importantly the first experiments at RHIC yield strong elliptic flow,
which, as it was mention in introduction, 
was not reproduced in any other model, except in fluid dynamical
models with a QGP EoS \cite{QM01}. This is a strong experimental indication
that transverse pressure builds up early in these reactions, in a few
$fm/c$, and strong stopping is also necessary to create strong flow before
Freeze Out, which usually happens when the radius of the system is not
more than $10\ fm$. 	 Our model builds up the locally equilibrated matter
- the initial conditions for further hydrodynamical expansion - at
$t_{lab}=3/5\ fm/c$ for central/peripheral regions at RHIC energies, which
is in agreement with previous estimations as well as the data.

The presentation in this section is based on
Refs. \cite{MCS00,CAM00,MCS01}, but we want to stress that the
preliminary results were presented at conferences and published in their
proceedings, while the model was still being developed and, thus, Ref.
\cite{MCS01} is more up to date than Refs. \cite{MCS00,CAM00}. And in
this work we also improved the presentation of our model, particularly in
sections \ref{four}, \ref{fvf}, \ref{ansol}. For the continuous description
we will repeat in this section some details from  Ref. \cite{MCS01}, but we
will also  underline the differences.

\subsection{Formulation of model}
\label{two}

The basic idea is to generalize the Bjorken model with baryon recoil,
presented in section \ref{bmw}, for collisions of two heavy ions and
improve it by strictly satisfying conservation laws
\cite{MCS00,CAM00,MCS01}. Similar ideas were also presented recently in
\cite{mishkap}.

First of all, we would like to create a grid in the $[x,y]$-plane ($z$ is
the beam axes, $[z,x]$ is the reaction plane). We will describe the
nucleus-nucleus collision as a sum of independent streak-by-streak
collisions corresponding to the same transverse coordinates, $\{x_i,
y_j\}$. We assume that baryon recoil for both target and projectile arise
from the acceleration of partons in an effective field, $F^{\mu\nu}$,
produced in the interaction. Of course, the physical picture behind this
model is based on chromoelectric flux tube or string models, but for our
purpose we consider $F^{\mu\nu}$ as an effective Abelian field. The most
important consequence of the non-Abelian fields, i.e., its self
interaction and the resulting flux tubes of constant cross section, are,
nevertheless, reflected in our model: we assume that the field is
one-dimensional. The fields generated by the colliding streaks are of
constant cross section during the whole evolution, and only their lengths
increase with time. As the string tension is constant, the energy of the
string increases linearly with increasing length.  The single
phenomenological parameter we use to describe this field must be fixed
from comparison with experimental data.

We describe the streak-streak collision based on the following set
of equations:
\beq
\partial_\mu \sum_i T_i^{\mu\nu}=\sum_i F_i^{\nu\mu} q_i N_{i \mu} \ ,
\eeq{eq1}
\beq
\partial_\mu \sum_i N_i^\mu = 0 \ , \quad i=1,2\ ,
\eeq{eq2}
where $N_i^\mu$ is the baryon current of $i$th nucleus, $q_i$ is the color
charge (discussed later in more detail). We are working in
the Center of Rapidity Frame (CRF), which is the same for all streaks.
We will use the parameterization:
\beq
N_i^\mu=n_i u_i^\mu \ ,
\quad
u_i^\mu=(\cosh y_i,\ \sinh y_i) \ .
\eeq{eq3}
$T^{\mu\nu}$ is the energy-momentum tensor. It
consists of five parts, corresponding to two nuclei and free field energy
(also divided into two parts) and one term defining the QCD perturbative
vacuum. (While in the similar equation in the Bjorken model with
baryon recoil, eq. (\ref{rec1}), $T^{\mu\nu}$ contains only the target
matter.)
$$
T^{\mu\nu}=\sum_{i=1,2} T_i^{\mu\nu}+T^{\mu\nu}_{pert}=
$$
\beq
\sum_{i=1,2}\left[ e_i\left(\left(1+c_0^2\right)u_i^\mu u_i^\nu
- c_0^2g^{\mu\nu}\right)
+T_{F,i}^{\mu\nu}\right]+B g^{\mu\nu}\ .
\eeq{eq5}
Here $B$ is the bag constant, the equation of state is $P_i=c_0^2 e_i$,
where $e_i$ and $P_i$ are the energy density and pressure of QGP.

Comparing our equation (\ref{eq1}) with the corresponding one in the
Bjorken model with baryon recoil, eq. (\ref{rec1}), one may notice that we
neglected the pion source term. As it was discussed in section \ref{bmw}
the source term takes care about the string hadronization process. We
are planning to use our model at the beginning of the ultra-relativistic
heavy ion collisions only, and the string decay process at later stages
will be described separately. Section \ref{four} is devoted to this
question. The advantage of equations (\ref{eq1}, \ref{eq2}) is a simple
analytic solution.

Within each streak we form only one flux tube with uniform field strength
or field tension, $\sigma$, between  the target and the projectile
streaks.  The field extends to the outside edges of the target and
projectile  streaks, i.e., the field also overlaps with the space-time
domains of the matter streaks. For practical and symmetry purposes we,
however, divide this field into two spatial domains, a target and a
projectile domain, ($i= 1,2$), separated at a fixed point,
$z_{sep}$, so that $\sigma_1 = \sigma_2 = \sigma$.  The choice of this
point will be specified later. (The field is constant and the only change
is that it extends with time at its two ends.)

In complete analogy to electro-magnetic field
\beq
F_i^{\mu\nu}=\partial^\mu A_i^{\nu}-\partial^\nu A_i^{\mu}=\left(
\begin{array}{cc}
0 & -\sigma_i \\
\sigma_i & 0
\end{array}\right) \ \ ,
\eeq{eq6}
\beq
\sigma_i=\partial^3 A_i^{0}-\partial^0 A_i^{3}\ ,
\eeq{eq7}
\beq
T_{F,i \mu\nu}=-g_{\mu\nu}\mathcal{ L}_{F,i}+\sum_\beta \frac{\partial \mathcal{ L}_{F,i}}
{\partial \left(\partial^\mu A_i^{\beta}\right)}\partial_\nu A_i^{\beta}
\ \ ,
\eeq{eq8}
\beq
\mathcal{ L}_{F,i} = - \frac{1}{4}F_{i \mu\nu}F_i^{\mu\nu}\ .
\eeq{eq9}

In our case the string tensions, $\sigma_i$, will be constant in the
space-time region after string creation and before string decay. The
creation of fields will be discussed later in more detail.

Since during the interpenetration both nuclei are closely following the
light  cone, the geometry of the reaction suggests to use the light cone
variables
\beq
(z,t)\rightarrow(x^+,x^-),\quad x^\pm=t\pm z \ ,
\eeq{eq12}
and to assume that $e_1, y_1, n_1$
are functions of $x^-$ only (the first nucleus goes from right to left,
$y_1(0)=-y_0<0$) and $e_2, y_2, n_2$ depend on $x^+$ only (the second
nucleus goes from left to right, $y_2(0)=y_0>0$).

In terms of lightcone variables (for the transformation matrixes between
space-time and lightcone coordinates see Ref. \cite{MCS01}):
\beq
N_i^\pm=N_{i, \mp}= n_i(u_i^0\pm u_i^3)= n_i e^{\pm y_i} \ ,
\eeq{eq13}
\beq
T_i=\left(
\begin{array}{cc}
T_i^{++} & T_i^{+-} \\
T_i^{-+} & T_i^{--}
\end{array}\right)=\frac{1}{2}
\left(
\begin{array}{cc}
h_{i+} e^{2 y_i} & h_{i-} \\
h_{i-} & h_{i+} e^{-2 y_i}
\end{array}\right) + T_{F,i}\ ,
\eeq{eq14}
where
\beq
h_{i+}=(1+c_0^2)e_i\ ,
\quad
h_{i-}=(1-c_0^2)e_i\ .
\eeq{eq16}
The other tensors in light cone variables are:
\beq
F_i=\left(
\begin{array}{cc}
F_i^{++} & F_i^{+-} \\
F_i^{-+} & F_i^{--}
\end{array}\right)=
\left(
\begin{array}{cc}
0 & \sigma_i \\
-\sigma_i & 0
\end{array}\right) \ .
\eeq{eq17}
\beq
T_{pert}=\left(
\begin{array}{cc}
0 & B \\
B & 0
\end{array}\right) \ .
\eeq{eq17a}
The energy-momentum tensor for free field in the light cone variables is:
\beq
T_{F,i}=\frac{1}{2}
\left(
\begin{array}{cc}
\sigma_i^2 & 0 \\
0 & \sigma_i^2
\end{array}\right) \ .
\eeq{eq22}

\begin{figure*}[htb]
\insertplotlong{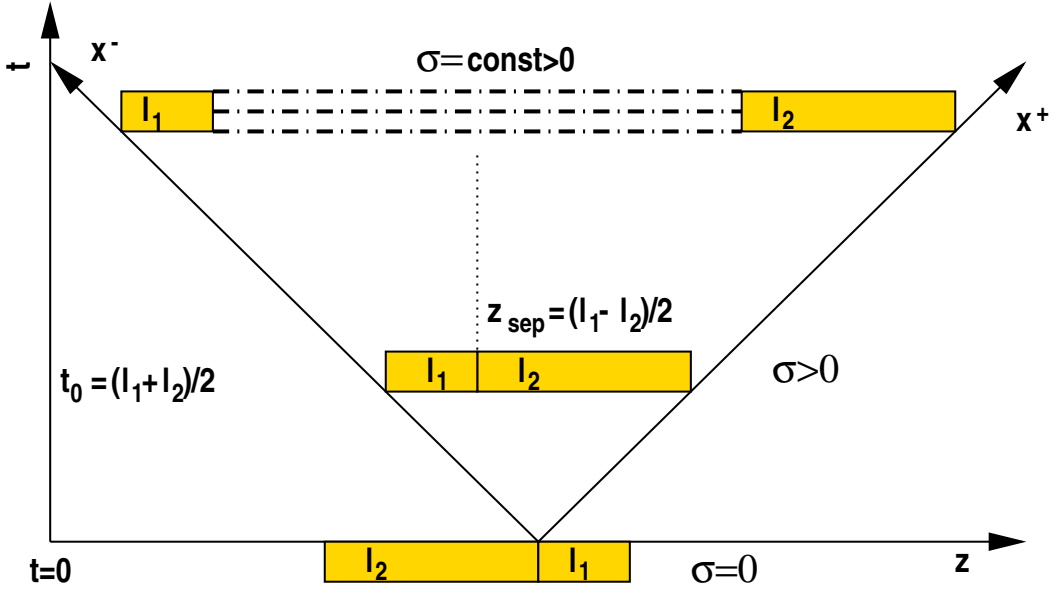}
\vspace{0.3cm}
\caption[]{Streak-streak collision. The time of first  touch of streaks is
$t=0$, and
$t=t_0$ corresponds to complete penetration of streaks through each other.
At this time strings are created, i.e., the string tension reaches an
absolute value $\sigma=A\left(\frac{\varepsilon_0}{m}\right)^2 n_0
\sqrt{l_1l_2}$ (\ref{eq44}). The dash-dotted lines denote the
chromoelectric field strings (it will be shown later that $\sigma$ is so
big that we can talk about several parallel strings or a ``string rope''). }

\label{fig1}
\end{figure*}

At the time of the first contact of the two streaks, $t=0$, there is no
string tension. We assume that strings are created, i.e., the string
tension achieves the value $\sigma$ at time $t=t_0$, corresponding to
complete penetration of streaks through each other (see Fig. \ref{fig1}).

\subsection{Conservation laws and string creation}

\label{three}

In lightcone variables eq. (\ref{eq2}) may be rewritten as
\beq
\partial_-N_1^-+\partial_+N_2^+=0 \ .
\eeq{eq19}
So, we have a sum of two terms, each depending on different independent
variables, and the solution can be found in the following way:
\beq
\begin{array}{ll}
\partial_-N_1^-=a, & \partial_+N_2^+=-a \ , \\
N_1^-=a x^- + (N_1^-)_0, & N_2^+=-a x^+ + (N_2^+)_0 \ ,
\end{array}
\eeq{eq20}
where the index $0$ indicates the initial proper density, which is the
normal nuclear density: $n_0=0.145\ fm^{-3}$. Since both $N_1^-$ and
$N_2^+$ are positive (and also more or less symmetric) we can conclude
that for our case $a=0$.

Finally,
\beq
N_1^-= n_1 e^{-y_1}= n_0 e^{y_0} \ , \quad
N_2^+= n_2 e^{y_2}= n_0 e^{y_0} \ ,
\eeq{eq21}
\beq
  n_1= n_0 e^{y_0+y_1}\ , \quad n_2= n_0 e^{y_0-y_2} \ .
\eeq{eq21a}
where $y_0\ (-y_0)$ is the initial rapidity of nucleus $2\ (1)$ in the
CRF, respectively. The other components are
given by eq. (\ref{eq13}).

Let us make an analogy to the electro-magnetic field, where two charges,
$q_1$  and $-q_2$,  move in the opposite directions, creating string-like
field between them, $\vec{E}=(0,0,E)$, which is constrained transversally
into a constant cross section. The Z-axis goes through the charges, $q_1$
and $-q_2$, and directed from $q_1$ to $-q_2$ (let us assume that we have
such a field configuration). So, forces acting on our charges, $q_1 E$ and
$-q_2 E$, have opposite sign  and both are working against the expansion
of the ``string''.  In our effective model we use color charges, and assume
that the vectors of these color charges point  in the opposite directions
in the color space \cite{ABP93}, so that the forces acting on both target
and projectile partons are opposite,  both stopping  the expansion of the
streak.  As our field strength (string tension, $\sigma$) is not yet
defined we normalize the charges to unity:
\beq
q_1=-q_2=1>0\ , \quad {\rm while} \quad \sigma_1=\sigma_2=\sigma\ .
\eeq{sch}
Then, we have the forces acting in $z$ direction: $q_1\sigma_1=\sigma\ ,
  $ and $\ q_2\sigma_2=-\sigma$.
Notice again that after string creation fields $\sigma_1(x)$ and
$\sigma_2(x)$  are spatially separated as well as the baryon densities,
$n_1$ and $n_2$; i.e., after complete penetration of  the initial streaks
through each other (see Fig. \ref{fig1}),
$\sigma_2$ acts on the  partons on the right side of the separating point
$z_{sep}=(l_1-l_2)/2$,    while the $\sigma_1$ acts on those on the left
side. In the absence of matter,  in the middle both fields are identical,
so the exact position of the  separating point does not play any role
until it does not enter  the target or projectile matter. The fields
$\sigma_i$  are generated by the corresponding four-potentials $A_i$,
which are different and spatially separated in the same way  (see the
dotted separating line, $z_{sep}=(l_1-l_2)/2$, in Fig. \ref{fig1}).

As described above we do not generate the chromoelectric field
self-consistently as a product of color currents, which are also affected
by the field. Our effective fields are external with respect to the
colliding  partons; that's why we can use the expression (\ref{eq22}) for
the field energy. On the other hand, if we want to satisfy the
conservation laws, we must  generate our effective fields in the collision
transferring energy from  matter to field. It's possible to define new
conserved quantities based on  eq. (\ref{eq1}).  Using the definition of
$F^{\mu\nu}$, eq. (\ref{eq7}), we can rewrite eq. (\ref{eq1}) as
$$
 \partial_\mu T^{\mu\nu} =\sum_i F_i^{\mu\nu} q_i N_{i, \mu}=
\sum_i q_i \left[\partial^\mu \left(A_i^\nu N_{i, \mu}\right)  -\right.
$$
\beq
\left.-A_i^\nu \partial^\mu N_{i, \mu}
-\partial^\nu \left(A_i^\mu N_{i, \mu}\right)
+A_i^\mu \partial^\nu N_{i, \mu}
\right] \ . 
\eeq{eq25}

The solutions for $N_1^-$ and $N_2^+$, eq. (\ref{eq21}), show that the
second term vanishes. The fourth term is a vector
$(A_1^-\partial_-N_1^+,-A_2^+\partial_+N_2^-)$ in lightcone coordinates.
So, if we impose the conditions
\beq A_1^-=0\ , \quad A_2^+=0
\eeq{apm} we can define a new energy-momentum tensor $\tilde{T}^{\mu\nu}$,
such that
\beq
\partial_\mu \tilde{T}^{\mu\nu}=0 \ ,
\eeq{eq26}
$$
\tilde{T}^{\mu\nu}=\sum_i \tilde{T_i}^{\mu\nu}+T^{\mu\nu}_{pert}=
$$
\beq
\sum_i \left(T_i^{\mu\nu} - q_i A_i^\nu N_i^\mu + g^{\mu\nu} q_i
A_i^\alpha N_{i \alpha} \right)+B g^{\mu\nu}
\eeq{eq27}

One may notice that the above defined new energy-momentum tensor is not
symmetric. We know that the energy-momentum tensor is not defined in
unique way - if $T^{\mu\nu}$ satisfies the eq. $\partial_\mu
T^{\mu\nu}=0$, then any other tensor $T^{\mu\nu}+\partial_\kappa
\Psi^{\mu\nu\kappa}$, satisfies it also, if $\Psi^{\mu\nu\kappa}=-
\Psi^{\nu\mu\kappa}$ \cite{lan1}. How to make it symmetric will be
discussed later in this section.

To satisfy the above choice of the fields, (\ref{sch}), and imposed
conditions, (\ref{apm}), we take the
vector potentials in the following form:
\beq
\begin{array}{ll}
A_1^- = 0, & A_1^+ = - \sigma_1 x^+ = - \sigma x^+, \\
A_2^- = \sigma_2 x^- = \sigma x^- ,& A_2^+ = 0 \ .
\end{array}
\eeq{eq24}
In principle the vector potentials are not defined in unique way -
the general expression would be:
\beq
\begin{array}{ll}
A_1^- = 0, & A_1^+ = - \sigma x^++a_1 x^-, \\
A_2^- = \sigma x^- +a_2 x^+,& A_2^+ = 0 \ .
\end{array}
\eeq{eq24g}
In our published work, Ref. \cite{MCS01}, the choice $a_1=a_2=0$  has
been used. It appears that this intuitive choice has a deep meaning -
with any other $a_1$ and $a_2$ the symmetrization procedure, which will be
discussed below, can not be applied.

How to define the $\sigma$ itself? In our
calculations we used the parameterization (see Refs. \cite{MCS00,CAM00,MCS01}):
\beq
\sigma=A\left(\frac{\varepsilon_0}{m}\right)^2 n_0
\sqrt{l_1l_2}=A\left(\frac{\varepsilon_0}{m}\right)
\frac{\sqrt{\sf n_1n_2}}{\dlt x \dlt y} \ ,
\eeq{eq44}
where $m$ is the nucleon mass, $\varepsilon_0$ is the
initial energy per nucleon,
$l_1$ and $l_2$ are the initial streak lengths in the CRF (see Fig.
\ref{fig1}), ${\sf n}_1$ and ${\sf n}_2$ are the numbers  of baryons in
the initial streaks, $\dlt x\dlt y$ is the cross section of the streaks,
and we've used the relation $\gamma=\varepsilon_0/m$.  Similar $({\sf n_1,
n_2})$-dependence of $\sigma=$ has been chosen in Ref. \cite{mishkap}. We
are working in the system  where $\ \hbar=c=1$, so $\sigma$ has a
dimension of $length^{-2}=energy/length$. The typical values of
dimensionless parameter $A$ are around $0.06-0.08$ (these values of $A$ are
for $\sigma$ measured already in $GeV/fm$). Notice, that there is only
one free parameter in parameterization (\ref{eq44}). The typical values of
$\sigma$ are $4-10\ GeV/fm$ for $\varepsilon_0=65 \ GeV$ per nucleon, and
$\sigma \approx 6-15\ GeV/fm$ for $\varepsilon_0=100 \ GeV$ per nucleon.
These values are
consistent with the energy density of the non-hadronized strings
in a reaction volume, or ``latent
energy density'', which is about $15\ GeV/fm^3$
for $\sqrt{s}=200\ GeV/nucl$ \cite{ASC91prl,ASC91pl,ACS92}.
Our $\sigma$ given by eq. (\ref{eq44}) grows linearly with
$\varepsilon_0$. This may be a too strong assumption, since we predict
much bigger field strength for LHC then the Monte Carlo string fusion
model \cite{ABP93}. It is easy to modify expression (\ref{eq44}) by
changing the power $2$ to $1+\alpha$, where $\alpha<1$, and probably this
will be done after the comparison with experimental data. In Ref. \cite{mishkap}
$\alpha=0.3$ has been chosen based on the low-x behavior of the nuclear
structure function or parton density \cite{seven}. Since such an
improvement is straightforward, we will keep it in mind, but for further
discussion we will use the definition (\ref{eq44}) as in Ref.
\cite{MCS00,CAM00,MCS01}.

Using the exact definition of $A_i^\mu$, eqs. (\ref{eq24}), eqs.
(\ref{eq14}, \ref{eq17a}, \ref{eq22}, \ref{sch}, \ref{eq27})  and
transformation matrixes from Ref. \cite{MCS01} we obtain
$$
\tilde{T}^{\mu\nu}=\left(
\begin{array}{cc}
\tilde{T}^{++} & \tilde{T}^{+-} \\
\tilde{T}^{-+} & \tilde{T}^{--}
\end{array}\right)=
\frac{1}{2}\left(
\begin{array}{cc}
h_{1+}e^{2y_1} & h_{1-} \\
h_{1-} & h_{1+}e^{-2y_1}
\end{array}\right)
$$
\beq
+\frac{1}{2}\left(
\begin{array}{cc}
h_{2+}e^{2y_2} & h_{2-} \\
h_{2-} & h_{2+}e^{-2y_2}
\end{array}\right)
+\frac{1}{2}\left(
\begin{array}{cc}
\sigma^2 & 2B \\
2B & \sigma^2
\end{array}\right)
\eeq{eq27a}
$$
+\left(
\begin{array}{cc}
-\sigma x^+ N_1^+ & 0\\
\sigma x^+ N_1^- & 0
\end{array}\right)
+\left(
\begin{array}{cc}
0&\sigma x^- N_2^+\\
0&-\sigma x^- N_2^-
\end{array}\right)
$$
Notice that the perturbative vacuum term, $B$, and the free field energy,
$\frac{\sigma^2}{2}$, cover all the interacting volume, while the
energy densities of matter and baryon currents are separated in space.

Now we are coming back to the question: how to symmetrize
$\tilde{T}^{\mu\nu}$? One may notice that the tensor
\beq
\tilde{T}_1^{\mu\nu}=\left(
\begin{array}{cc}
0&\sigma x^- N_2^+\\
\sigma x^+ N_1^-&0
\end{array}\right)\ ,
\eeq{symm}
satisfies the equation $\partial_\mu \tilde{T}_1^{\mu\nu}=0$ (it's 
easy to check
using the eqs. (\ref{eq21})). Thus we can redefine $\tilde{T}^{\mu\nu}$ in a
symmetric way:
\beq
\tilde{T}^{\mu\nu}:=\tilde{T}^{\mu\nu}-\tilde{T}_1^{\mu\nu}\ .
\eeq{tmnsym}
This important step has not been done in Refs. \cite{MCS00,CAM00,MCS01}.
It causes some  changes in the calculations of the energy densities at
$t=t_0$ (see \ref{app41}), but we will see that this does not affect
appreciably the final results.

Now the new conserved quantities are
\beq
Q^0=\int \tilde{T}^{00} dV = \dlt x\dlt y
\int \tilde{T}^{00} dz
\ ,
\eeq{eq28}
\beq
Q^3=\int \tilde{T}^{03} dV = \dlt x \dlt y
\int \tilde{T}^{03} dz
\ .
\eeq{eq29}

We can rewrite the energy-momentum tensor in $(t,z)$ coordinates:
$$
\tilde{T}^{\mu\nu}=\left(
\begin{array}{cc}
\tilde{T}^{00} & \tilde{T}^{03} \\
\tilde{T}^{30} & \tilde{T}^{33}
\end{array}\right)=
\left(\begin{array}{cc}
\frac{\sigma^2}{2}+B & 0 \\
0& \frac{\sigma^2}{2}-B
\end{array}\right)+
$$
$$
\left(
\begin{array}{cc}
(e_1+P_1)\cosh^2 y_1 - P_1& (e_1+P_1)\cosh y_1 \sinh y_1\\
(e_1+P_1)\cosh y_1 \sinh y_1& (e_1+P_1)\sinh^2 y_1 + P_1
\end{array}\right)+
$$
\beq
\left(
\begin{array}{cc}
(e_2+P_2)\cosh^2 y_2 - P_2& (e_2+P_2)\cosh y_2 \sinh y_2\\
(e_2+P_2)\cosh y_2 \sinh y_1& (e_2+P_2)\sinh^2 y_2 + P_2
\end{array}\right)
\eeq{eq27d}
$$
-\frac{\sigma x^+}{2}\left(
\begin{array}{cc}
  N_1^+ & N_1^+ \\
N_1^+ & N_1^+
\end{array}\right)
-\frac{\sigma x^-}{2}\left(
\begin{array}{cc}
  N_2^- & - N_2^- \\
- N_2^- & N_2^-
\end{array}\right).
$$
Obviously, in the  absence of the fields, before string creation and after
string decay, $\tilde{T}^{\mu\nu}\equiv T^{\mu\nu}$ and, thus, $(Q^0,Q^3)$
reverts to $(P^0,P^3)$ - components of the four-momenta of  the system.

Based on conservation of $Q^0,\ Q^3$ we can calculate energy densities,
$e_1(t_0),\ e_2(t_0)$, at the moment $t=t_0$, when the string with tension
$\sigma$ is created - eqs. (\ref{eq42a}, \ref{eq42b}). Another result
from \ref{app41} we will need  for later discussion is the rapidity
of the CM frame:
\beq
\tanh y_{CM} = Q^3/Q^0 = v_0/M \ ,
\eeq{cmrap} where $M=\frac{l_2+l_1}{l_2-l_1}\ .$

Thus, for $x^\pm>x_0$ we should solve eqs. (\ref{eq26}),
with boundary conditions
\beq
\begin{array}{ll}
N_1^\pm (x^-=x_0)= n_0 e^{\mp y_0} &
N_2^\pm (x^+=x_0)= n_0 e^{\pm y_0} \\
h_{1+}(x^-=x_0)=e_1(t_0)(1+c_0^2) &
h_{2+} (x^+=x_0)=e_2(t_0)(1+c_0^2) \\
y_1(x^-=x_0)=-y_0 & y_2(x^+=x_0)=y_0 \\
\sigma_1(x^-=x_0)=\sigma & \sigma_2(x^+=x_0)=\sigma\\
q_1(x^-=x_0)=1 & q_2(x^+=x_0)=-1\ ,
\end{array}
\eeq{eq45}
where $x_0=2t_0-|z(0)|$ defines the string creation surface $t=t_0$,
for a parton or cell element in the position $z=z(0)$ at the time $t=0$.
The complete analytical solution can be found in Ref. \cite{MCS01}.

\subsection{Hadronization of the String Ropes}

\label{four}

\begin{figure}[htb]
\insertplots{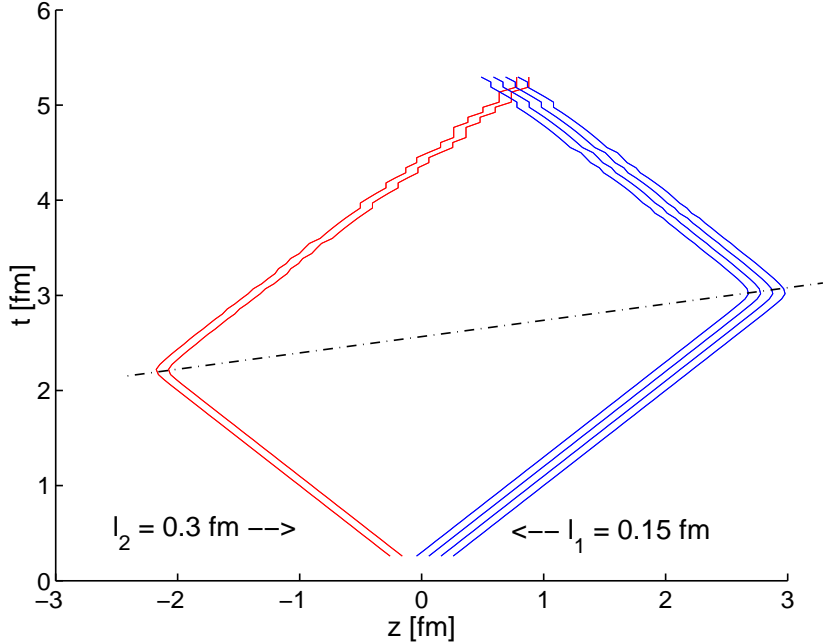}
\caption[]{The typical trajectory of the initial streaks with lengths
$l_1$ (coming from the right)  and $l_2$ (coming from the left).}
\label{traj0}
\end{figure}

If we let the partons (or cell domains) evolve according to
trajectories derived in our model (see eqs. (37, 38) in Ref. \cite{MCS01}),
the partons from the colliding initial streaks
will keep going in the initial direction, gradually slowing down up to
some time $t=t_{i,turn}$; then they will turn and go backwards until the
two streaks again penetrate through each other (see Fig. \ref{traj0}).
Of course, it is hard to believe that such a contraction of the reaction
volume could
really happen in heavy ion collisions, because of string decays,
string-string interaction, interaction between streaks and other reasons,
which would be very difficult to take into account.
As we already pointed out
earlier we did not install any string hadronization mechanism in our model
and, thus, there is no way to decrease the energy supplied in the strings.
If we try to describe the evolution of the system
with our model after one or both
initial streaks turn back, we run into the trouble, since the length
of the string starts to decrease, so the field energy should also
decrease, but it has no chance to do this.
Therefore the above simple analytical model
fails to describe the Yo-Yo motion for the initial streaks - it has the
realistic character until the length of the final streak has started
to decrease, i.e., approximately up to the dash-dotted line in Fig.
\ref{traj0}.

In Refs. \cite{MCS00,CAM00,MCS01} we assumed that the final result of the
collision of two streaks, after stopping the string's expansion and after
its decay, is one streak of the length $\dlt l_f$ with homogeneous energy
density distribution, $e_f$, moving like one object with rapidity
$y_f$. We assumed that this is due to string-string interactions and string
decays, which we are not going to describe in our simple model.
As mentioned above the typical values of the string
tension, $\sigma$, are of the order of $10\ GeV/fm$, and these may be
treated as several parallel strings. The string-string interaction will
produce a kind of ``string rope'' between our two streaks, which is
responsible for final energy density distribution.

One of the simplest ways to quantitatively take into account string 
decays is presented
in Ref. \cite{mishkap}. One of the complications appearing in this case is 
that the hadronization of the strings has to be described in the proper time  
(i.e. all quantities became functions of $(\tau, \eta),\ \tau=\sqrt{t^2-z^2}$)
which is, in general, different for all the pairs of colliding streaks. Then 
in order to be able to transform all of the final streaks to one frame, say
CRF, we would have to know the details of the their dynamics for a long 
period of the proper time, $t=\sqrt{\tau^2+z^2}$. Unfortunately 
this is not the case in real modeling, for example in Ref. \cite{mishkap}
authors do not describe the dynamics after all the chromoelectric 
strings for the particular streak pair decay.  

The real situation may be more complicated: when the energy accumulated in
the strong color fields will be finally released in a production of
$q\bar{q}$ pairs and gluons, this process may noticeably change the
composition of matter as compared to the chemical equilibrium case
\cite{KM85}. Therefore, the matter created after the mutual stopping
of interpenetrating streaks can not, in general, be described
by the equilibrium EoS.

For the stopped and equilibrated matter a set of homogeneous final streaks
is the simplest assumption. Its advantages
are: simple expressions for $e_f,\ y_f$ and a very simple way to generate
an initial state  for further hydrodynamical description. We have to give
only four numbers (coordinates  of the ends of the final streak, $z_{left}$
and $z_{right}$, $e_f$ and $y_f$) for  each point of a transverse grid,
$\{x_i, y_j\}$ . Later in section \ref{ansol} we  present an analytical
solution for the expanding final streaks which seems to be more realistic.
Remember - we describe the initial state which is not directly observable
in experiments. Thus, even a flat initial rapidity distribution may end
up after the hydrodynamical evolution in both a forward-backward peaked or
a centrally peaked distribution depending on several circumstances.

\begin{figure*}[htb]
\insertplotlong{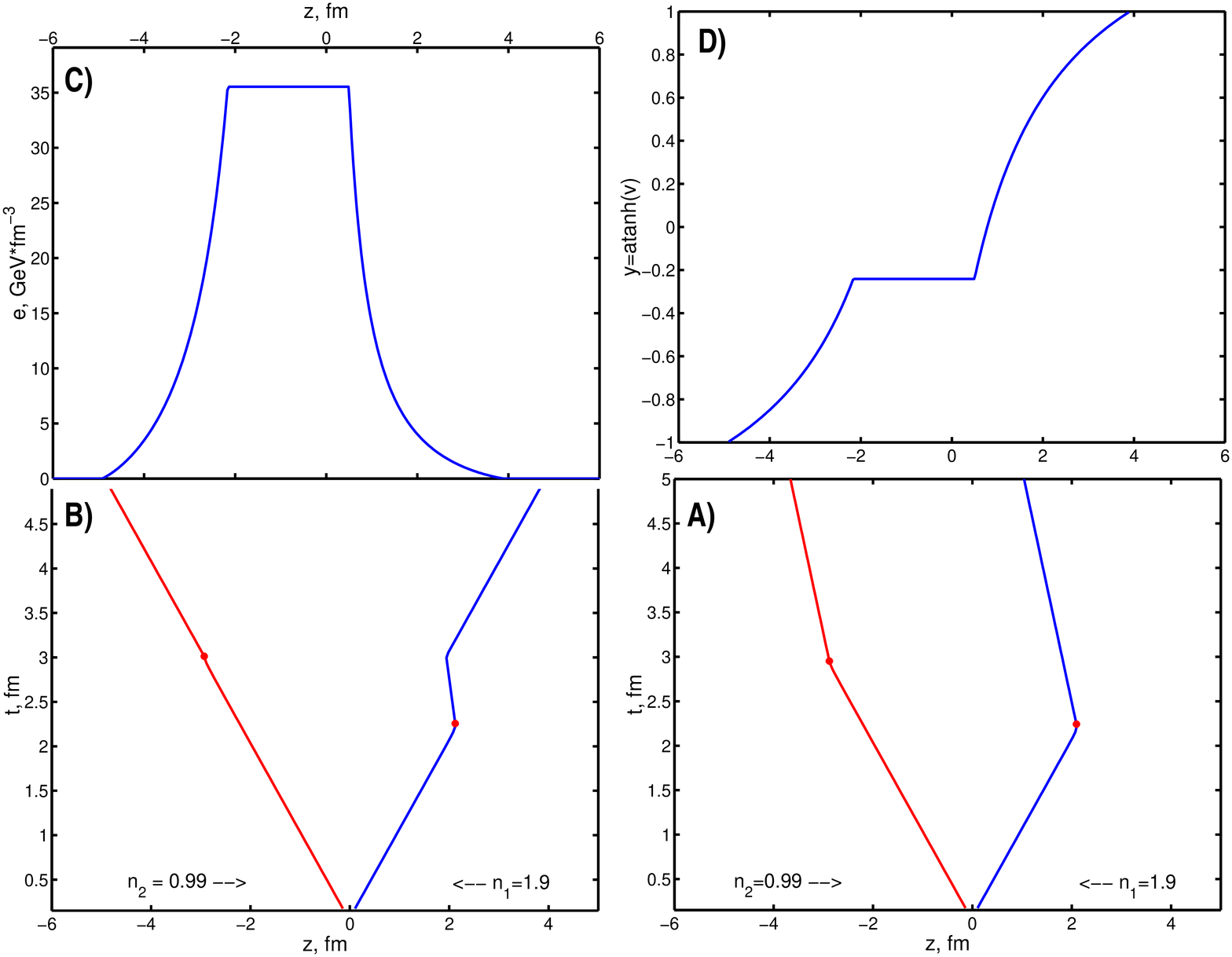}
\caption[]{A) The typical trajectory of the ends of two initial streaks
corresponding to the numbers of nucleons $n_1$ and $n_2$,
$\varepsilon_0=65\ GeV/nucl$, $A=0.09$ (the parameter $A$ was introduced
in  eq. (\ref{eq44})).  Stars denote the points where $y_i=y_f$. From
$t=t_0$ until these stars, the streak ends move according trajectories
derived  in Ref. \cite{MCS01} (see eqs. (37, 38)). Then the final streak
starts to move  like a single object with rapidity $y_f$, eq. (\ref{eq66}),
in CRF.
\\ B) The same situation as in subplot (A), but for expanding final
streak. C) Shows $e(z)$ profiles of expanding final streak 
($t_h=5 \ fm$). We can clearly
see  three regions - two of forward and backward rarefaction waves and a
middle where the initial energy density,
$e_f$ (eq. (\ref{eq69})), is still preserved.
D) Shows the rapidity profile of the expanding final streak ($t_h=5 \ fm$). 
We can clearly
see  three regions - two of forward and backward rarefaction waves and a
middle where the initial rapidity,
$y_f$ (eq. (\ref{eq66})), is still preserved. }
\label{fig2}
\end{figure*}

The final energy density and rapidity, $e_f$ and
$y_f$, should be determined from conservation laws.  The conservation of
the energy and momentum gives for the  final rapidity:
\beq
\cosh^2 y_{f} = \begin{array}{c}
(M^2(1+c_0^2)-2c_0^2v_0^2)+\\
\underline{\sqrt{(M^2(1+c_0^2)-2c_0^2v_0^2)^2+4c_0^4v_0^2(M^2-v_0^2)}}\\
2(1+c_0^2)(M^2-v_0^2) \end{array}\ ,
\eeq{eq66}
where we neglected $B \dlt l_f$ next to $Q^0/\dlt x\dlt y$
and use notation $M$, introduced in eq. (\ref{Mnot}); $v_0=\tanh y_0$
is the initial velocity\footnote{Please notice the misprint in equation
for $\cosh^2 y_{f}$ in Refs. \cite{MCS00,CAM00,MCS01}:
$(M^2(1+c_0^2)+2c_0^2v_0^2)$ instead of $(M^2(1+c_0^2)-2c_0^2v_0^2)$.}.
Then the expressions for $e_f$ is:
\beq
e_f=\frac{{Q^0\over\dlt x \dlt y}}{((1+c_0^2)\cosh^2 y_f
- c_0^2)
\dlt l_f} \ .
\eeq{eq69}

The typical trajectory of the streak ends is presented in Fig. \ref{fig2}
(A). From $t=t_0$ they move according to trajectories derived in Ref.
\cite{MCS01} (see eqs. (37, 38)) until they reach the rapidity
$y_i=y_f$. Later, the final streak starts to move like one object with
uniform rapidity, $y_f$, until we reach the time  when the fluid dynamical
calculation starts. The time and position of final streak formation can be
found from the condition $y_i=y_f$ (see eq. (43) in Ref. \cite{MCS01}). We
shall call this scenario the ``nonexpanding final streak assumption''.

Unfortunately this simple model of the final state causes a principle
problem. Comparing expressions (\ref{eq66}) and (\ref{cmrap}) one may
notice a strange thing: the final streak moves like one object with a
certain rapidity calculated from  energy-momentum conservation, but this
rapidity, $y_f$, may differ, in general, from the rapidity of the Center of
Mass of the system given by  eq. (\ref{cmrap}). Obviously, if all the
elements of the system move with the same rapidity, then the Center of Mass
of the system also moves with the same rapidity. This problem have been
pointed out in the Refs. \cite{MCS00,CAM00,MCS01}, but the above
oversimplified  assumptions have been kept because of the simple
analytical solution for the $e_f$, $y_f$. We think that now we have a
deeper understanding of the source of the problem and we would like to
update the discussion in Ref. \cite{MCS00,CAM00,MCS01}. This is
illustrated and elucidated in next section.


\subsection{Flow velocity field generated by pressure in relativistic
fluid mechanics}
\label{fvf}

\begin{figure}[htb]
	\insertplots{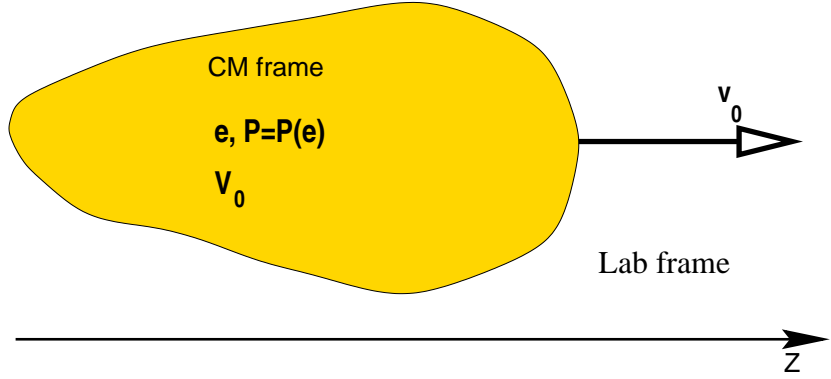}
\caption[]{The homogeneous peace of matter
with energy density $e$ and
EoS $P=P(e)$ moving with velocity $v_0$ in the Lab frame.
}
\label{f_task_a}
\end{figure}

Let us propose to the reader the following problem:

{\it Imagine the homogeneous piece of matter with energy density $e$ and
Equation of State $P=P(e)$ moving with velocity $v_0$ in the Lab frame
(see Fig. \ref{f_task_a}). The question is to find the energy-momentum
tensor, $T^{\mu\nu}$, and four-momentum,
$P^\mu$, in the Center of Mass frame and in the  Lab frame and show that
they are consistent with each other.  }


\paragraph{CM frame:} \
In the CM frame, which is in our case the rest frame of matter the energy
momentum tensor and four-momentum of the system are:
\beq
T^{\mu\nu}=\left(
\begin{array}{cccc}
e & 0 & 0 & 0 \\
0 & P & 0 & 0 \\
0 & 0 & P & 0 \\
0 & 0 & 0 & P
\end{array}\right)\ ,
\eeq{cm1}
\beq
P^\mu=\left(
E,\ 0,\ 0,\ 0
\right)\ ,
\eeq{cm2}
where $E=eV_0$, $V_0$ is the volume of our matter in the CM frame, $\mu,\
\nu=0,1,2,3$.

The four-momentum is connected to the energy-momentum tensor in the
following way
\cite{lan1}:
\beq
P^\mu=\int T^{\mu\nu} dS_\nu \ ,
\eeq{pT0}
where the integral is over any 3-dimensional hypersurface covering all the
volume, i.e., intersecting all the world lines of the matter in
space-time. If we integrate over the hyperplane $x^0=t=const$, then
\beq
P^\mu=\int T^{\mu0} dV \ .
\eeq{pT}
It's easy to check that equations (\ref{cm1}) and (\ref{cm2}) are indeed
connected by the relation (\ref{pT}) for any EoS.

\paragraph{Lab frame:} \
In the Lab frame our system is moving with velocity $\vec{v}=(0,0,v_0)$.
The four-momentum undergoes the Lorentz transformation:
\beq
P_L^\mu=\left(
E\gamma_0,\ 0, \ 0,\ E\gamma_0 v_0
\right)\ .
\eeq{lab2}

The invariant expression for the energy-momentum tensor in any reference
frame is
  \beq
T^{\mu\nu}=(e+P)u^\mu u^\nu-Pg^{\mu\nu}\ .
\eeq{tmn}

In our case $u^\mu=\gamma_0(1,0,0,v_0)$. Thus,
\beq
T_L^{\mu\nu}=\left(
\begin{array}{cccc}
(e+P)\gamma^2_0-P & 0 & 0 & (e+P)\gamma^2_0 v_0\\
0 & P & 0 & 0 \\
0 & 0 & P & 0 \\
(e+P)\gamma^2_0 v_0& 0 & 0 & (e+P)\gamma^2_0 v^2_0+P
\end{array}\right)\ ,
\eeq{lab1}
Now we have to use
eq. (\ref{pT}) to find $P_L^\mu$ from $T_L^{\mu\nu}$.
\beq
P_L^0=\int dV \left[(e+P)\gamma^2_0-P\right]=
\left[(e+P)\gamma^2_0-P\right] V_0/\gamma_0\ ,
\eeq{labe}
\beq
P_L^3=\int dV (e+P)\gamma^2_0 v_0=
(e+P)\gamma^2_0 v_0 V_0/\gamma_0 \ ,
\eeq{labp}
where $V_0/\gamma_0$ is the volume of our system in Lab frame.
Comparing eqs. (\ref{labe}, \ref{labp}) with eq. (\ref{lab2}) we obtain
the following system of equations:
\beq
\left\{\begin{array}{l}
\frac{(e+P)\gamma^2_0-P}{\gamma_0}= e \gamma_0 \\
(e+P)\gamma_0 v_0= e \gamma_0 v_0
\end{array}\right.
\eeq{sys0}
and we
see that for a consistent description we have to put $P=0$.
Why do we get such a strong restriction on our EoS?!

\begin{figure}[htb]
	\insertplots{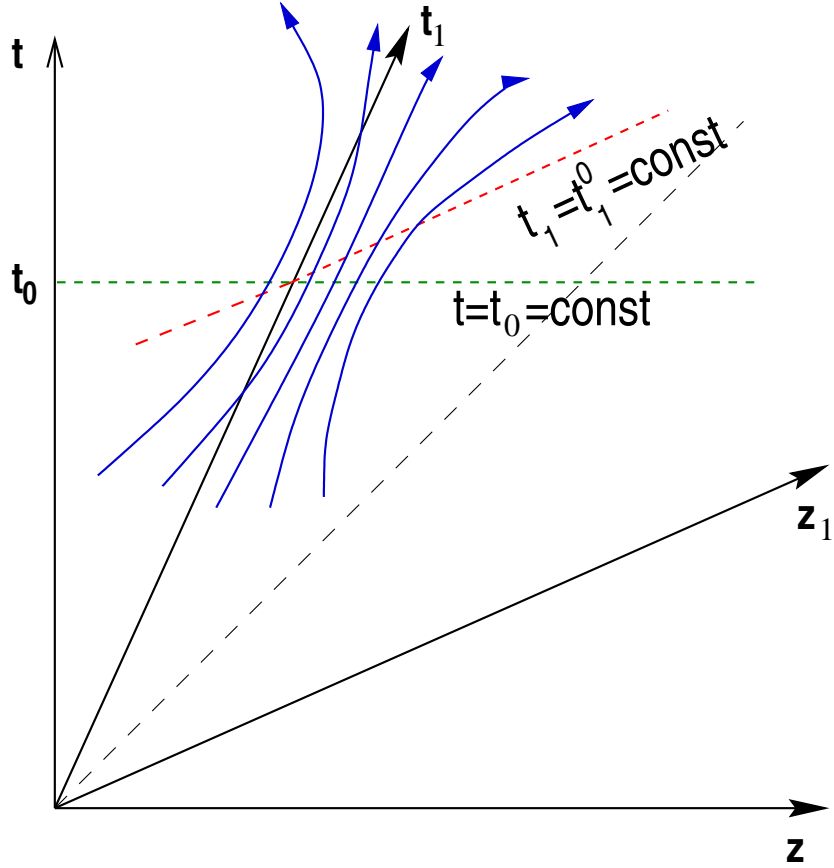}
\caption[]{All the matter on
the hypersurface $t=t_0$ in the Lab frame has the same velocity, but in the
CM frame on the hypersurface $t_1=t_1^0$ different parts of matter
have different velocities.}
\label{parall}
\end{figure}

Certainly the problem is that we made wrong assumptions. We assumed
that all the parts of our system are moving with the same velocity. This is
true for the dust, where the pressure is exactly zero. If $P>0$ such a
situation may happen for some particular $t$ in the particular reference
frame (remember we integrate $T^{\mu\nu}$ over the hypersurface
$t=const$). If we boost our system the parts of matter on the new
integration hypersurface $t_1=const$ will not move with the same
velocity, as illustrated in Fig. \ref{parall}: in the Lab frame all
the matter on the hypersurface $t=t_0$ has the same velocity, but in the
CM frame we have to either integrate over hypersurface $t_1=t_1^0$, where
different parts of matter have different velocities or integrate over
hypersurface $t=t_0$ homogeneous in velocity, but in this case $dS_\mu
\ne (dV, \vec{0})$ in eq. (\ref{pT0}).  Briefly speaking the presence of
pressure requires, in general, the flow velocity field in CM frame,
$\vec{v}_{CM}(t,\vec{r})$, for real physical situation, and our invariant
energy-momentum tensor, $T^{\mu\nu}$, defined by eq. (\ref{tmn}) seems to
be clever enough to know that.

Thus, as we see, the assumption of the final streak moving like one
object oversimplifies the real physical situation. The way to improve the
model is to let the final streak expand with the rapidity field which
will allow us to satisfy exactly all the conservation laws. We can
describe such a streak expansion based on the analytical solution for
one--dimensional expansion of matter into the vacuum.

\subsection{Expanding final streaks}
\label{ansol}
Apart of the principal reasons discussed above, our test results 
\cite{CAMS02,CAMS01}
indicated that the uniform final streaks without expansion
lead to somewhat unrealistic expansion patterns. Thus we had to improve
the model and include the expansion of the final streaks.

In this section we discuss the one-dimensional expansion of the
finite streak into the vacuum (generalizing the description in \cite{R95}).
The initial condition is
\begin{eqnarray} \label{in1}
e (z,t_0) & = & \left\{ \begin{array}{ll}
          0&~~-\infty <z < R_-~,\\
		 e_0 &,~~R_-\leq z \leq R_+ \\
              0 &,~~R_+ <z < \infty~,
               \end{array} \right. \\
v(z,t_0) & = & \left\{ \begin{array}{ll}
              -1 &,~~-\infty < z< R_- \\
              \tanh y_0 &,~~R_- \leq z \leq R_+ \\
              1 &,~~R_+ <z < \infty~,
               \end{array} \right. \label{in2}\ ,
\end{eqnarray}
where $R_-$, $R_+$ are the borders of the system.
The choice $v=\pm 1$ in the vacuum is purely conventional, but it
guarantees a continuous hydrodynamic solution at the boundary to the
vacuum, since in the limit of infinite dilution the velocity of matter
approaches unity.

Now consider a simple wave, i.e., continuous hydrodynamic flow.
This flow respects entropy conservation \cite{LL6}.
For isentropic flow, the $t-$ and $z-$component of the equation of motion,
$\partial _{\mu }T^{\mu \nu } = 0$, can be combined to yield
\begin{equation}
\left(\frac{\partial}{\partial t} + \frac{ v \pm c_0 }{ 1 \pm v c_0}
\frac{\partial}{\partial z} \right) \mathcal{ R}_{\pm} = 0~,
\end{equation}
where the so-called {\em Riemann invariants\/} are
\begin{equation} \label{Rpm}
\mathcal{ R}_{\pm} \equiv y - y_0 \pm \int_{e_0}^e \frac{ c_0 \,
{\rm d}e' }{e' + p(e')}~.
\end{equation}
Obviously, the Riemann invariants are constant,
\begin{equation}
\left. \frac{{\rm d} \mathcal{ R}_{\pm}}{{\rm d} t} \right|_{\mathcal{ C}_{\pm}}=0~,
\label{ch2}
\end{equation}
along the so-called characteristic curves $\mathcal{ C}_{\pm}(x,t)$,
the positions $x_{\pm}(t)$ of which are determined by integrating
\begin{equation} \label{ch1}
\frac{{\rm d}x_{\pm}}{{\rm d}t} = \frac{ v \pm c_0 }{ 1 \pm
v c_0}~.
\end{equation}
The characteristics are the world-lines of sonic disturbances
on the hydrodynamic flow pattern. For simple waves moving to the right
(i.e., $v >0$),
one can prove \cite{courant} that $\mathcal{ R}_+ = const$, and it
suffices to consider the $\mathcal{ C}_-$--characteristic. For simple
waves moving to the left ($v<0$), $\mathcal{ R}_-=const$, and only
the $\mathcal{ C}_+$--characteristics have to be considered.

For the initial conditions (\ref{in1}, \ref{in2}) we determine the value
of $\mathcal{ R}_\pm$ at $t=t_0,\ z=R_\pm$ to be $\mathcal{ R}_\pm=0$ and thus
\begin{equation}
y_\pm =y_0 \mp \int_{e_0}^e \frac{ c_0\,{\rm d}e' }{e' +
p(e')}~.
\end{equation}
If we consider the QGP EoS,
$p(e) = c_0^2\, e~,~~c_0^2 = const$,
this gives
$$
v(e)_\pm = \tanh y(e)_pm =
\tanh \left[ y_0\mp \frac{c_0}{1+c_0^2} \ln \left\{ \frac{e}{e_0}
\right\} \right]= 
$$
\beq
=\frac{\tanh y_0 \mp \frac{1-(e/e_0)^{2c_0/(1+c_0^2)}}{
1+(e/e_0)^{2c_0/(1+c_0^2)}}}{1 \mp \tanh y_0
\frac{1-(e/e_0)^{2c_0/(1+c_0^2)}}{
1+(e/e_0)^{2c_0/(1+c_0^2)}}}~.
\eeq{vp}
This equation nicely shows the defining
property of a simple wave, namely that there is a unique
relationship between the value of the fluid velocity and its
thermodynamic state \cite{courant}. Furthermore, with the help of
(\ref{ch1}) for $\mathcal{ C}_\mp$, we can now calculate for any given $e$,
$0 \leq e \leq e_0$, the position $z(t;e)$
at which this value for the energy density occurs for given $t$,
\begin{equation} \label{x}
z_\pm(t;e)-R_\mp = \frac{v(e) \pm c_0}{ 1 \pm v(e)~c_0}~(t-t_0)~.
\end{equation}
Eq.\ (\ref{x}) has
similarity form, i.e., the profile of the rarefaction wave
does not change with time when plotted as a function of the
similarity variable $\zeta_\pm \equiv \frac{z\mp-R_\pm}{t-t_0}$.

Let us now complete the solution of the hydrodynamic problem. Causality
requires that the initial conditions (\ref{in1}, \ref{in2}) remain
unchanged for $|\zeta_\pm| > 1$. Thus, we only have to determine the
solution in the range $-1 \leq \zeta_\pm \leq 1$, i.e. in the forward
light cone (remember that $\zeta_+$ ($\zeta_-$), obtained from the $\mathcal{
C}_-$  ($\mathcal{ C}_+$)--characteristic, eq. (\ref{x}), corresponds to
$y_+$ ($y_-$), obtained from $\mathcal{ R}_+=const$ ($\mathcal{ R}_-=const$)
condition).

For the forward going rarefaction wave, $v>0$, generated on the right end
of the initial streak,  from eqs.\ (\ref{vp}, \ref{x}) we infer that the
head of this wave (the point where the rarefaction of matter starts, i.e.,
where the energy density $e$ starts to fall below $e_0$) travels with the
velocity $\frac{\tanh y_0-c_0}{1-c_0\tanh y_0}$ to the left. On the other
hand, the base of the rarefaction wave (the point where the vacuum ends,
i.e., where $e$ starts to acquire non-vanishing values) travels with light
velocity $v = 1$ to the right. Thus, the energy density as a function of
$\zeta$ can be written as
\begin{equation}
e(\zeta_+) = e_0 \cdot \left\{ \begin{array}{l}
          1~, \\
	  ~~~~~-1 \leq \zeta_+ \leq \frac{\tanh y_0-c_0}{1-c_0\tanh y_0} \\
      {\displaystyle \left[\frac{1+\tanh y_0}{1-\tanh y_0}~ \frac{1-c_0}{1+c_0}~
	 \frac{1-\zeta_+}{1+\zeta_+}
           \right]}^{(1+c_0^2)/2c_0}~, \\
	   ~~~~~  \frac{\tanh y_0-c_0}{1-c_0\tanh y_0}< \zeta_+ \leq 1~.
                \end{array} \right.
\label{epl}
\end{equation}
The velocity can then be inferred from (\ref{vp}), or simply
from (\ref{x}):
\beq
\tanh y(\zeta_+) = \left\{ \begin{array}{ll}
           \tanh y_0 &,~~-1 \leq \zeta_+ \leq \frac{\tanh 
y_0-c_0}{1-c_0\tanh y_0} \\
      {\displaystyle \frac{c_0+\zeta_+}{1+\zeta_+ c_0}}&,~~
		  \frac{\tanh y_0-c_0}{1-c_0\tanh y_0}< \zeta_+ \leq 1~.
                \end{array} \right.
\eeq{ypl}

The similar analyzes for the backward going rarefaction wave, $v<0$,
generated on the left end of the initial streak, gives:
\begin{equation}
e(\zeta_-) = e_0 \cdot \left\{ \begin{array}{l}
           1~,\\
	   ~~~~~\frac{\tanh y_0+c_0}{1+c_0\tanh y_0} \leq \zeta_- \leq 1 \\
      {\displaystyle \left[\frac{1-\tanh y_0}{1+\tanh y_0}~ \frac{1-c_0}{1+c_0}~
	 \frac{1+\zeta_-}{1-\zeta_-}
           \right]}^{(1+c_0^2)/2c_0}~,\\
	   ~~~~~  -1< \zeta_- \leq \frac{\tanh y_0+c_0}{1+c_0\tanh y_0}~,
                \end{array} \right.
\label{emin}
\end{equation}
\beq
\tanh y(\zeta_-) = \left\{ \begin{array}{ll}
           \tanh y_0 &,~~ \frac{\tanh y_0+c_0}{1+c_0\tanh y_0} \leq 
\zeta_- \leq 1\\
      {\displaystyle \frac{\zeta_--c_0}{1-\zeta_- c_0}}&,~~
		  -1< \zeta_- \leq \frac{\tanh y_0+c_0}{1+c_0\tanh y_0}~.
                \end{array} \right.
\eeq{ymin}

This simple analytical solution is valid as long as the two rarefaction
waves  did not overlap in the middle  of the system. Further evolution
becomes more complicated and doesn't have a similarity form any more.

The point where two rarefaction waves start to overlap, $z_{over},\
t_{over}$ can be found from the system
\beq
\left\{ \begin{array}{l}
  \zeta_+(z,t)=\frac{z-R_+}{t-t_0}=\frac{\tanh y_0-c_0}{1-c_0\tanh y_0} \\
  \zeta_-(z,t)=\frac{z-R_-}{t-t_0}=\frac{\tanh y_0+c_0}{1+c_0\tanh y_0}\ ,
		  \end{array} \right.
\eeq{over}
which gives
\beq
z_{over}=\begin{array}{c}
R_+(1-\tanh y_0 c_0)(\tanh y_0 + c_0)\\
\underline{-R_-(1+\tanh y_0 
c_0)(\tanh y_0 - c_0)}\\
2c_0(1-\tanh^2 y_0)
\end{array}\ ,
\eeq{zover}
\beq
t_{over}=t_0+\frac{(z_{over}-R_+)(1-\tanh y_0 c_0)}{(\tanh y_0 - c_0)}\ .
\eeq{tover}
Thus the analytical solution (\ref{epl}-\ref{ymin}) valid for $t_0 
\leq t \leq t_{over}$.

For the symmetric initial state, $R_+=-R_-=R$, eqs. 
(\ref{zover},\ref{tover}) became
\beq
z_{over}=R\frac{\tanh y_0 (1-c_0^2)}{c_0(1-\tanh^2 y_0)}\ ,
\eeq{zover1}
\beq
t_{over}=t_0+R\frac{1-\tanh y_0 c_0}{c_0(1-\tanh^2 y_0)}\ .
\eeq{tover1}

Thus, based on the solution presented above we can have a more advanced
description of the final streaks than the one presented in section
\ref{four}. Let us assume that the homogenous final streaks, with some
$e_f$, $y_f$,  are formed in the CRF when the larger of initial streaks
reach the rapidity $y_f$. Up to this point the fluid cell trajectories
are the same as those discussed in section \ref{four}, but after the
homogeneous final streak is formed, it starts to expand according to
the simple rarefaction wave solutions presented above. Figure \ref{fig2}
(B) shows new trajectories of the streak ends, and Figures \ref{fig2}
(C,D) present energy density and rapidity profile for this expanding
streak.

Such an initial state with expanding streaks will also help to avoid the
problem which may cause the development of numerical artifacts, namely
a step-like in beam direction initial energy density distribution (output
of ESRM): it has a jump from $e(x_i,y_j)=const$ inside the matter to $0$
in the outside vacuum (of course, where is also a jump of $e$ as a
function of $x,y$, but it is much smoother and this is not the direction of
the initial expansion). In order  to avoid (or at least to suppress) the
effect of this, it was proposed in Ref. \cite{hirano} to smooth over
initial energy  density distribution, for example by a Gaussian shape.
Our simple analytic  solution smooths over this jump in a natural
way.

\subsection{Initial conditions for hydrodynamical
calculations}

\label{inc}

\begin{figure}[htb]
\insertplots{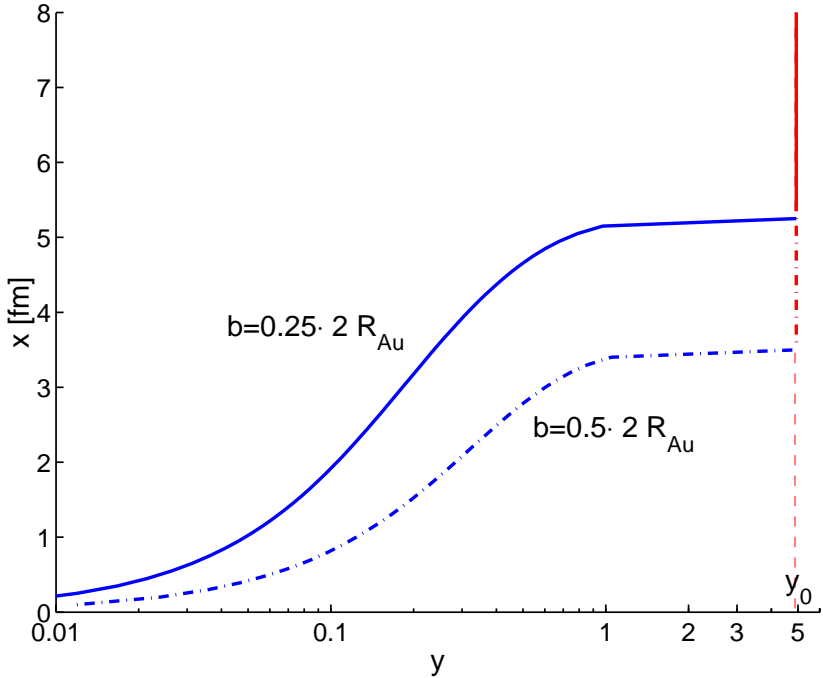}
\caption[]{Final rapidity, $y_f$, profiles of the final streaks in the
reaction plane
for Au+Au collision at
$\varepsilon_0=65\ GeV/nucl$,
$A=0.08$.
The rapidities of the final streaks in CRF are calculated according to eq.
(\ref{eq66}). Our profiles are in agreement with the schematic sketch in
paper \cite{SSVWX} and with the scenario  discussed in ``firestreak'' model
(see Fig. \ref{fs}). }

\label{rap65}
\end{figure}

In this section we present the results of our calculations.  We are
interested in the shape of the QGP formed when the expansion of the
combined target-projectile streaks stop and their matter is locally
equilibrated. This will be the initial state for further hydrodynamical
calculations. The time, $t_h$, at which we start the hydrodynamical
description is a second (in addition to $A$) free parameter of our model.
Of course, $t_h$ should be larger than the time of final streak formation,
at least in the central, most hot and dense region. For the peripheral
streaks the string tension is low, and the transparency is large, but
peripheral matter does not play a leading role in further hydrodynamic
expansion. Therefore, we will also build the final streaks ($y_f,\ e_f$)
for peripheral streak-streak collisions, with lengths, $\dlt l_f$,
corresponding to the lengths of the interacting region at the moment
$t=t_h$, even if the final rapidity,
$y_f$, was not yet achieved for this particular collision.\footnote{ The
calculations of the initial state presented in this section are done for
colliding nuclei with a homogeneous density distribution. The modification
is pretty straightforward - we can use the Wood-Saxon density
distribution for the Au nucleus in our case. This will lead to the
increasing of the density in the middle of the colliding nuclei, therefore
the middle steaks will contain more partons, will generate a stronger field
and, thus, will form the final streaks faster. While the peripheral
streaks will contain less matter, therefore their dynamics, which is
oversimplified, will have less influence on the final state.} On the
other hand the time, $t_h$, cannot be longer than the time for which the
simple analytical solution presented in the previous section is valid.
I.e., $t_h \leq Minimum\{t_{form,i}+t_{over,i}\}$, where $t_{form,i}$ is a
time of formation of the final streak for particular streak-streak
collisions $i$: when the larger of two initial streak gets rapidity $y_f$
(given by eq. (43) in Ref. \cite{MCS01}); $t_{over,i}$ is given by eq.
(\ref{tover}).  In this section we present all the results at $t_h=5\
fm/c$, while for
$\varepsilon_0=100\ GeV/nucl$,
$b=0.25\cdot2\ R_{Au}$, $A=0.08$ the maximal possible
$t_h = Minimum\{t_{form,i}+t_{over,i}\}\approx 5.04\ fm/c$.

The reader may notice that all of the above discussion about final streaks
is for the energy density distribution. What about the baryon charge
distribution in the final streaks?  Our knowledge here is very poor and, in
principle, we are free to assume anything between two extreme cases: A) the
complete transparency picture, where charges would remain at the ends of
the final streak; this is the qualitative picture in Ref. \cite{mishkap};
B) homogeneous in baryon charge final streaks, the  simplified situation
discussed in Refs.
\cite{MCS00,CAM00,MCS01}

\begin{figure*}[htb]
\insertplotlong{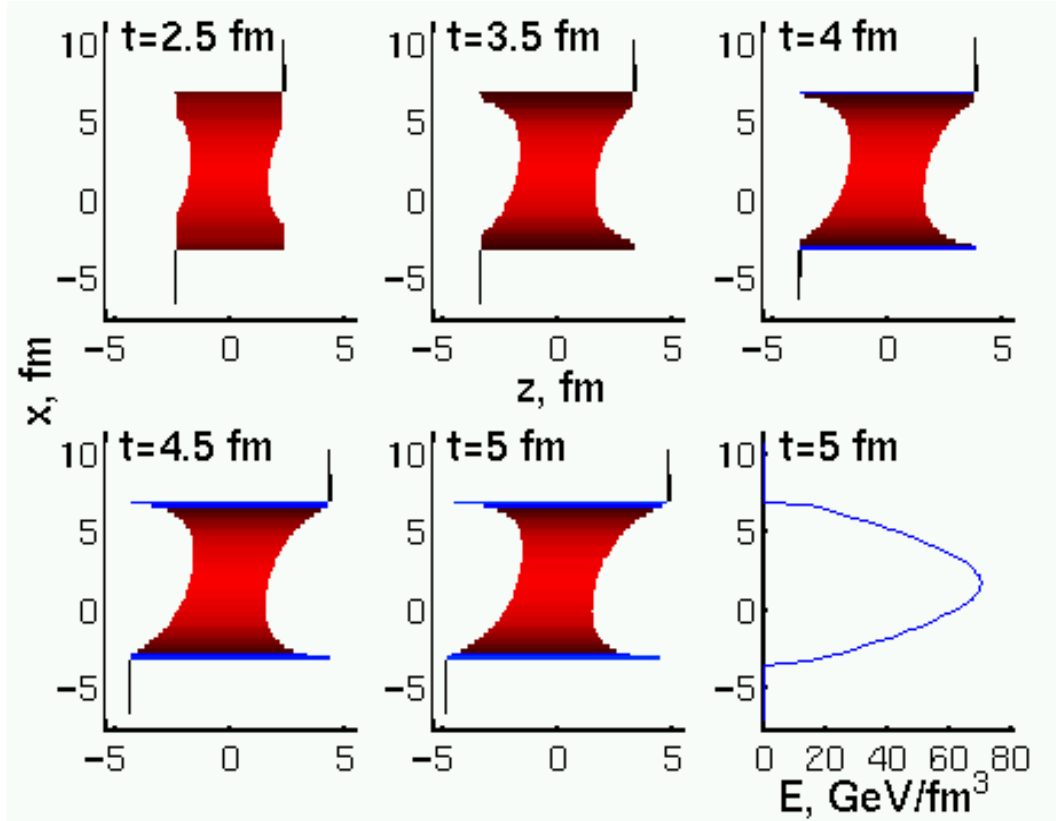}
\caption[]{Au+Au collision at $\varepsilon_0=65\ GeV/nucl$,
$b=0.25\cdot2\ R_{Au}$, $A=0.08$
(the parameter $A$ was introduced in eq. (\ref{eq44})),
$E=T^{00}$ is presented in the reaction plane as a
function of $x$ and $z$ for different times in the laboratory frame. The
final shape of the QGP volume is a tilted disk $\approx 45^0$, and the
direction of the fastest expansion will deviate from both the beam axis
and the usual transverse flow direction and will generate the
third flow component \cite{Dr-th01,CAMS02,CAMS01}. Note that the initial state
for hydro looks pretty much like the one discussed in the ``firestreak''
model (see Fig. \ref{fs}), since the final streaks are moving with
rapidities presented in Fig. \ref{rap65}. These calculations are done for
the symmetrized $\tilde{T}^{\mu\nu}$ (see eq. \ref{tmnsym}), but the
results do not differ too much from what is presented in Ref.
\cite{MCS01}. }
\label{ev11}
\end{figure*}

\begin{figure*}[htb]
\insertplotllarge{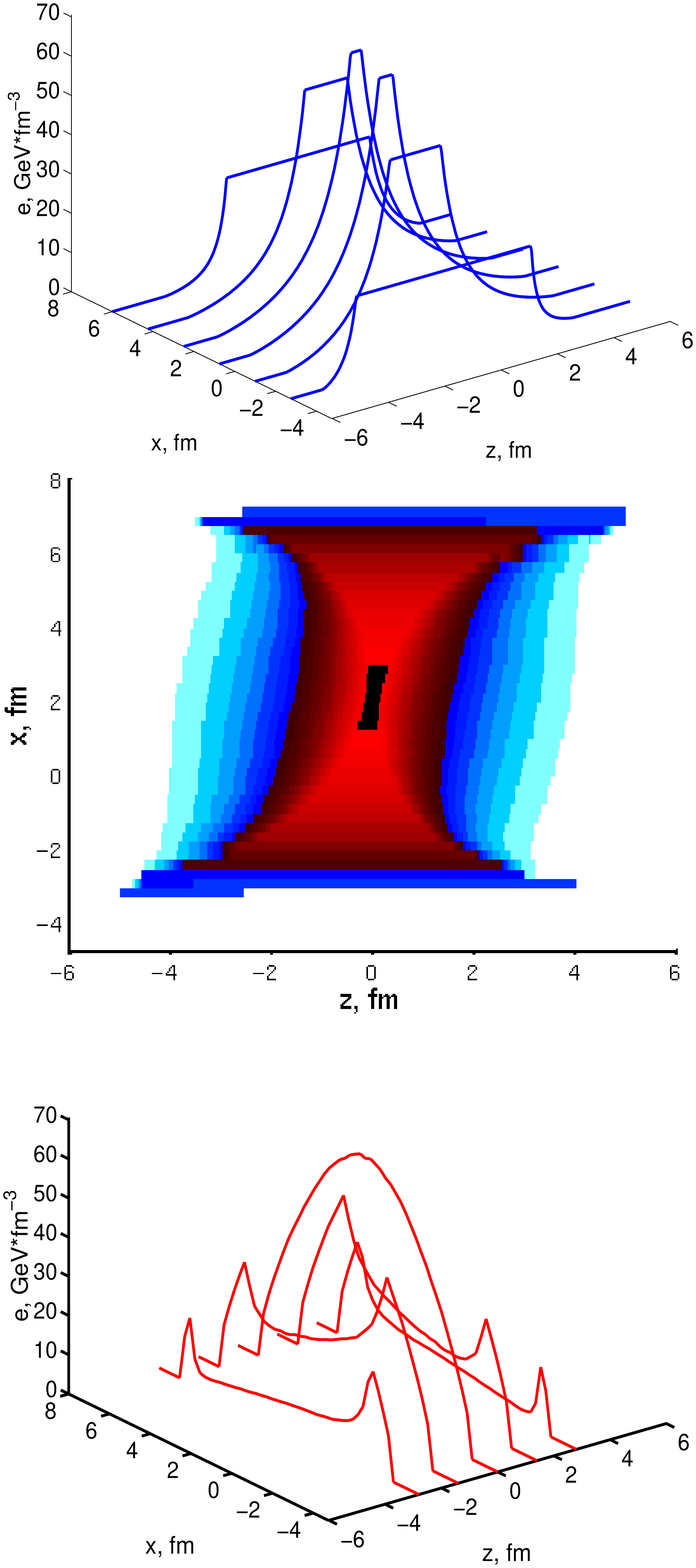}
\caption[]{Simulation of the same collisions as in Fig. \ref{ev11}
(Au+Au, $\varepsilon_0=65\ GeV/nucl$, $b=0.25\cdot2\ R_{Au}$, $A=0.08$),
but with expanding final streaks.
The middle subplot presents $e(x,z)$ in the reaction plane for $t_h=5\ fm/c$.
The upper subplot shows the $e$ profiles for different $x=const$. 
We can clearly see 
three regions - two of
forward and backward rarefaction and middle where the initial energy density,
$e_f$ (eq. (\ref{eq69})), is still preserved.
The lower subplot shows the $e$ profiles for different $z=const$.
}
\label{xzprof}
\end{figure*}

First, we present the initial rapidities for final streaks for different
$x$ coordinates in the reaction plane -- Fig. \ref{rap65}. Fig. \ref{ev11} shows
the energy density distribution in the reaction plane with non-expanding
final streaks (Au+Au collision at $\varepsilon_0=65\ GeV/nucl$,
$b=0.25\cdot2\ R_{Au}$, $A=0.08$). Fig. \ref{xzprof} shows the energy
density distribution in the reaction plane for the same collision, but with
expanding streaks. The QGP forms a tilted disk for
$b\not =0$. Thus, the direction of fastest expansion, the same as the
largest pressure gradient, will be in the reaction plane, but will deviate
from both the beam axis and the usual transverse flow direction. So, the
new flow component, called ``antiflow'' or ``third flow component'' (see
section \ref{3flow}), appears in addition to the usual transverse flow
component in the reaction plane (for non-expanding streaks this was shown
in \cite{Dr-th01,CAMS02,CAMS01}). With increasing beam energy the usual
transverse flow is getting weaker, while this new flow component is
strengthened.

\begin{figure*}[htb]
\vspace{-0.5cm}
\insertplotllargee{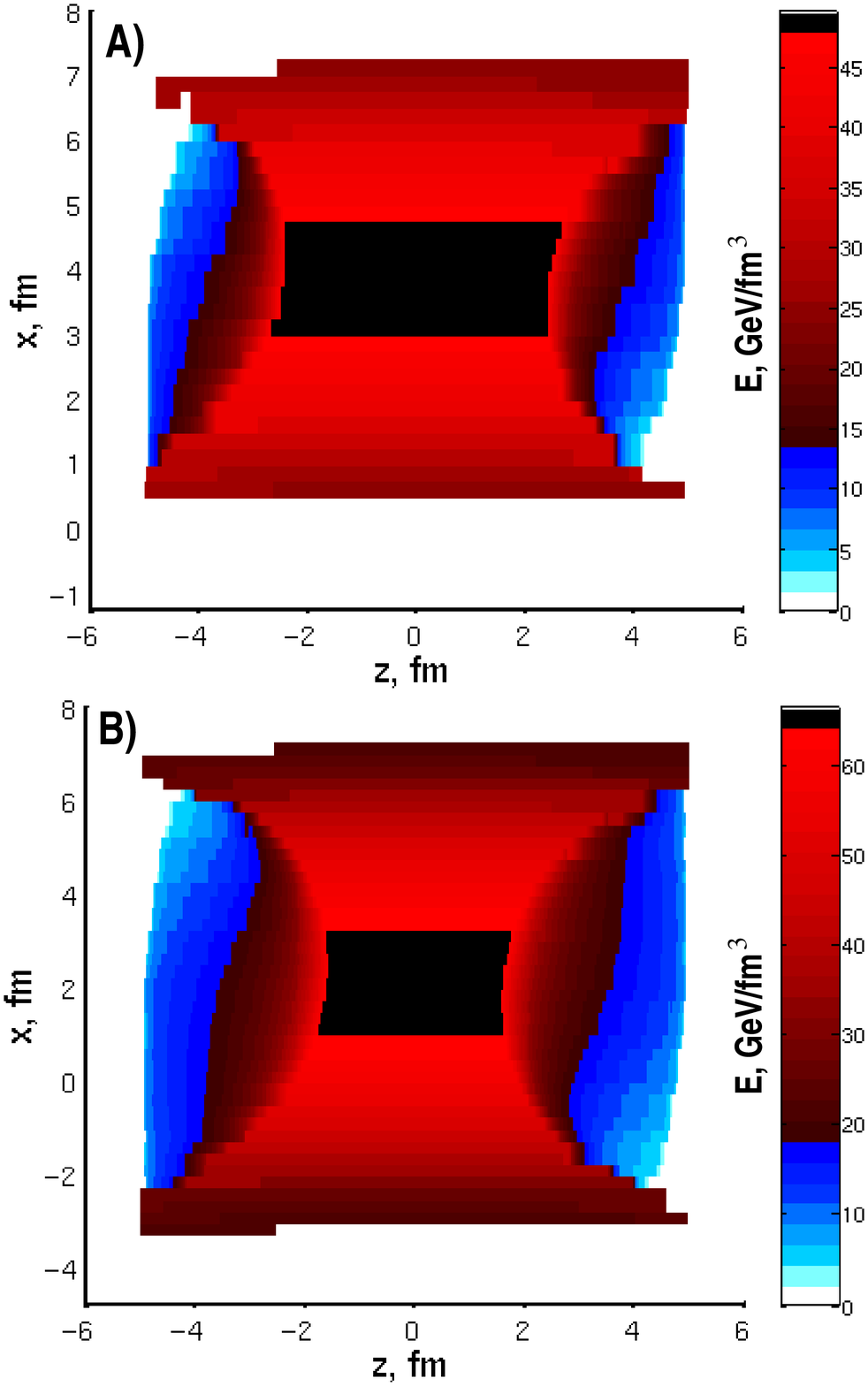}
\caption[]{Au+Au collision at $\varepsilon_0=100\ GeV/nucl$,
$A=0.065$ ,
$E=T^{00}$ is presented in the reaction plane
as a function of $x$ and $z$ for $t_h=5\ fm/c$.
Subplot A) $\left(b=0.5\cdot2\ R_{Au}\right)$,
subplot B) $\left(b=0.25\cdot2\ R_{Au}\right)$. The QGP volume has a
shape of a tilted disk
and may produce a third flow component. }
\label{a065}
\end{figure*}

\begin{figure*}[htb]
\vspace{-0.5cm}
\insertplotlargee{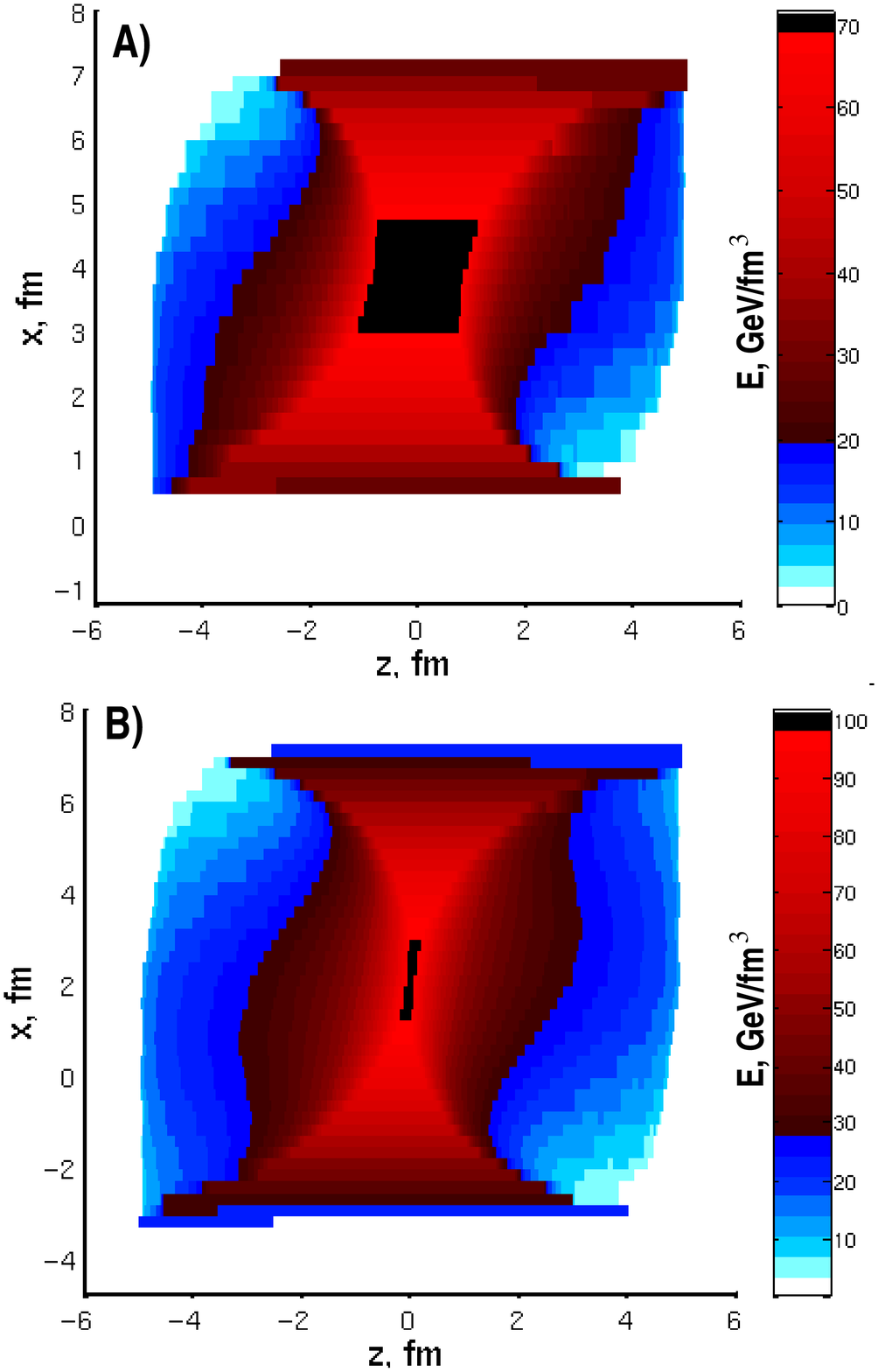}
\caption[]{The same as Fig. \ref{a065}, but for $A=0.08$.
We see that due to a larger string tension the energy density is much
larger. }
\label{a08}
\end{figure*}

More results for full RHIC energy ($\varepsilon_0=100\ GeV/nucl$) for
different string tensions and impact parameters are presented in Figs.
\ref{a065}--\ref{a08}.

Note that the initial state with non-expanding streaks (Figs. \ref{ev11})
looks pretty much like the one discussed in ``firestreak'' model (see Fig.
\ref{fs}), since the final streaks are moving with rapidities presented
in Fig. \ref{rap65}. If we let final streaks expand, this smooths over
picture, but most of the matter, nevertheless, keeps similar energy
density profile and velocity distributions.

 From Figs. \ref{ev11} and \ref{a065}-\ref{a08} we may see that for the
central collisions at $\varepsilon_0=100\ GeV/nucl$ the maximum energy
density $E=T^{00}$  reaches as high as $E_{max}\approx 100\ GeV/fm^3$. It
seems to be larger than what one should expect from the Bjorken estimate
(see section \ref{bm}) - $e_0=4.3\ GeV/fm^3$ for
$\sqrt{s}=130\, A\ GeV$ \cite{milov}. This is natural since our initial
energy density distribution is much more peaked as a function of $z$ than
is the one obtained in the Bjorken picture (our energy density distribution
is also peaked as a function of transverse coordinate $(x,y)$, but here
the difference seems to be not that large, see for example \cite{EK01-p}).
One should, nevertheless, keep in mind that we have also a  huge initial
pressure, $P=c_0^2 e$, which will cause a fast  hydrodynamical expansion,
and thus, as discussed in section \ref{bm}, will reduce
$\left(\frac{d E_{T}}{d y}\right)_{y=0}$. This is a usual situation  for
hydrodynamical calculation, for example, in Ref. \cite{EK01-p}
$e_0=23\ GeV/fm^3$ has been used to reproduce the elliptic flow at RHIC
for $\varepsilon_0=65\ GeV/nucl$,  although the Bjorken initial geometry
had been assumed.

In a recent work \cite{larisa} another scenario of the creation of the
third flow component based on the shadowing picture was proposed. The
calculations were performed in the framework of QGSM. The shadowing effect
manifests itself in the following way - hadrons emitted at small
rapidities in  the antiflow direction can propagate freely, while hadrons
emitted  in the normal directed flow direction still remain within the
expanding subsystem of  interacting particles.    We would like to point
out how one can distinguish this scenario from the  one discussed above in
this paper. The tilted initial state has the maximal effect for  non-zero,
but not very large, impact parameters, approximately up to
$b=0.5(R_A+R_B)$, i.e., in semicentral reactions. For semiperipheral and
peripheral collisions our string rope tension, $\sigma$,  would be small,
and we would have to wait for a long time for the final streak formation,
i.e., at the initial stages we would have one-dimensional  Bjorken
transparency. On the other hand, shadowing just becomes  stronger with
increasing impact parameter and gives the most drastic effect for
semiperipheral and peripheral collisions.


\section{Conclusions}

\label{ch-7}
A realistic and detailed description of an energetic heavy ion reaction
requires a Multi Module Model, wherein the different stages of the
reaction are each described with a suitable theoretical approach. The
initial stages are the most problematic at RHIC energies. We tried
to show that for the  qualitative understanding of the basic features of
the initial stages a phenomenological ``firestreak'' model can be used.
For the detailed calculations of the initial state the  Effective String
Rope model has been developed. This model is  based on earlier Coherent
Yang-Mills field theoretical models and introduces an effective string
tension based on Monte-Carlo string cascade and parton cascade model
results. Contrary to earlier expectations based on standard string
tensions of $1\ GeV/fm$, which lead to the Bjorken model type of initial
state, the effective string tension introduced in our model  for
collisions of energetic massive heavy ions  causes limited transparency.
The increased string tension is a consequence of collective effects
related to QGP formation. These collective effects in central and 
semicentral collisions lead to an effective string tension of the order of
$10\ GeV/fm$ and consequently cause much less transparency than earlier
estimates. The resulting initial locally equilibrated state of matter in
semicentral collisions takes a rather unusual form (see Figs.
\ref{ev11},
\ref{xzprof},
\ref{a065}-\ref{a08}) and the resulting flow velocity field  (see Fig.
\ref{rap65}, \ref{fig2} (D)) is  in agreement with the  ``firestreak'' phenomenological
picture. Such an  initial state will manifest itself by the asymmetry of
the produced collective flow, even after the subsequent hydrodynamical
expansion \cite{Dr-th01,CAMS02,CAMS01}.

\section*{Acknowledgment}

Two of us (V.K.M and D.D.S.) thank the  support of the Bergen
Computational Physics Laboratory in the framework of the European
Community - Access to Research Infrastructure  action of the Improving
Human Potential Programme and  the Humboldt Foundation.

\appendix

\section{Initial conditions after string creation}
\label{app41}

Our conserved quantities are (\ref{eq28}, \ref{eq29})
\beq
Q^0=\int \tilde{T}^{00} dV = \dlt x\dlt y
\int \tilde{T}^{00} dz
\ ,
\eeq{eq28a}
\beq
Q^3=\int \tilde{T}^{03} dV = \dlt x \dlt y
\int \tilde{T}^{03} dz
\ ,
\eeq{eq29a}
where $\tilde{T}^{00}$ and $\tilde{T}^{03}$ are given by eq. (\ref{eq27d}).
Before string creation the initial values of the modified
energy-momentum tensor,
$\tilde{T}^{\mu\nu}$, are
\beq
\tilde{T}_1^{00}=\tilde{T}_2^{00}=e_0\cosh^2 y_0=
\left(\frac{\varepsilon_0}{m}\right)^2 n_0 m \ ,
\eeq{eq34.0}
\beq
\tilde{T}_2^{03}=-\tilde{T}_1^{03}=e_0\tanh y_0 \cosh^2 y_0=
\left(\frac{\varepsilon_0}{m}\right)^2 n_0 m v_0\,
\eeq{eq34.3}
where $m$ is the nucleon mass, $\varepsilon_0$ is the initial
energy per nucleon, and
we have used
$\cosh^2 y_0=\gamma_0^2=\left(\frac{\varepsilon_0}{m}\right)^2$.
$v_0=\tanh y_0$
is the initial velocity, $v_0=1$ is a good approximation for
ultra-relativistic heavy ion collisions.
Thus,
\beq
Q^0=\dlt x \dlt y \left(\frac{\varepsilon_0}{m}\right)^2 n_0 m
(l_1+l_2) \ ,
\eeq{eq32}
\beq
Q^3=\dlt x \dlt y \left(\frac{\varepsilon_0}{m}\right)^2 n_0 m
(l_2-l_1) v_0 \ ,
\eeq{eq33}
where $l_1$ and $l_2$ are the initial lengths
of streaks (see Fig. \ref{fig1}), $\dlt x$, $\dlt y$ are the grid sizes
in $x$ and $y$ directions.
Thus the rapidity of CM frame is
\beq
\tanh y = \frac{Q^3}{Q^0} = \frac{v_0}{M} \ ,
\eeq{cmrap0}
where
\beq
M=\frac{l_2+l_1}{l_2-l_1}\ .
\eeq{Mnot}

After string creation --
$$
\tilde{T}^{00}=e_1 \cosh^2 y_1 +c_0^2e_1 \sinh^2 y_1
+e_2 \cosh^2 y_2 +c_0^2e_2\sinh^2 y_2
$$
\beq
+{1\over 2}\sigma^2+B-
\frac{\sigma x^+}{2} n_0 e^{-y_0}
e^{2y_1}
-\frac{\sigma x^-}{2} n_0 e^{-y_0}
e^{-2y_2} \ ,
\eeq{eq36}
$$
\tilde{T}^{03}=e_1(1+c_0^2) \cosh y_1\sinh y_1
+e_2(1+c_0^2) \cosh y_2\sinh y_2
$$
\beq
-\frac{\sigma x^+}{2} n_0 e^{-y_0}
e^{2y_1}
+\frac{\sigma x^-}{2} n_0 e^{-y_0}
e^{-2y_2} \ ,
\eeq{eq38}

At the point of complete penetration of streaks, $t=t_0=(l_1+l_2)/2$
(see Fig \ref{fig1}), we introduced energy densities $e_1(t_0)$ and
$e_2(t_0)$.
We assumed transparency, i.e., that complete penetration happened so fast,
that the field itself that was created during this time did not have time
to stop partons. So, the rapidities $y_{1(2)}(t_0)=-y_0(y_0)$
correspondingly, and  the proper baryon densities did not change, and
thus, the baryon current  is conserved.
Then the energy and momentum conservation laws can be written in the form:
$$
\frac{Q^0}{\dlt x \dlt y}=
\left[(1+c_0^2) \cosh^2 y_0 - c_0^2\right]\left(e_1(t_0)l_1+
e_2(t_0)l_2\right)
$$
\beq
+\left(\frac{\sigma^2}{2}+B\right)(l_1+l_2)
+\frac{\sigma n_0 e^{-y_0}}{4}\left(l_1^2+l_2^2\right)\ , 
\eeq{eq40}
$$
\frac{Q^3}{\dlt x \dlt y}=
\left[(1+c_0^2) \cosh^2 y_0\right]\left(-e_1(t_0)l_1+
e_2(t_0)l_2\right)
$$
\beq
-\frac{\sigma n_0 e^{-y_0}}{4}\left(l_1^2-l_2^2\right)
\ .
\eeq{eq41}

We neglect $c_0^2$ next to $(1+c_0^2) \cosh^2 y_0$ in eq. (\ref{eq40}),
then eqs. (\ref{eq40}, \ref{eq41}) may be solved
\beq
e_1(t_0)=
\frac{n_0 m}{1{+}c_0^2}-
\frac{{\sigma^2\over 2} + B}{ \left(\frac{\varepsilon_0}{m}\right)^2
  (1{+}c_0^2)}
\frac{l_1{+}l_2}{2l_1}-
\frac{\sigma n_0 e^{-y_0}}{4\left(\frac{\varepsilon_0}{m}\right)^2
(1{+}c_0^2)}l_1\ ,
\eeq{eq42a}
\beq
e_2(t_0)=
\frac{n_0 m}{1{+}c_0^2}-
\frac{{\sigma^2\over 2} + B}{ \left(\frac{\varepsilon_0}{m}\right)^2
  (1{+}c_0^2)}
\frac{l_1{+}l_2}{2l_2}-
\frac{\sigma n_0 e^{-y_0}}{4\left(\frac{\varepsilon_0}{m}\right)^2
(1{+}c_0^2)}l_2\ .
\eeq{eq42b}

\end{document}